\documentclass[a4,11pt]{article} 
\usepackage{amssymb,amsthm,url,paracol}
\usepackage{graphicx,hyperref,url}
\usepackage{threeparttable}
\usepackage[english]{babel}
\usepackage{authblk}
\usepackage[a4paper, margin=1in]{geometry}
\hypersetup{
    colorlinks = true,
    linkcolor = blue,
    anchorcolor = blue,
    citecolor = blue,
    filecolor = blue,
    urlcolor = blue
    }    
\renewcommand*{\Affilfont}{\footnotesize}

\title{The power of relativistic jets: a comparative study}

\author[1]{Luigi Foschini\footnote{Email: {\tt luigi.foschini@inaf.it}; Tel.: +39-02-72320-458.}}
\author[2,1]{Benedetta Dalla Barba}
\author[3]{Merja Tornikoski}
\author[4]{Heinz Andernach}
\author[5]{Paola Marziani}
\author[6]{Alan P. Marscher}
\author[6]{Svetlana G. Jorstad}
\author[7]{Emilia J\"arvel\"a}
\author[8]{Sonia Ant\'on}
\author[9]{Elena Dalla~Bont\`a}

\affil[1]{\Affilfont Osservatorio Astronomico di Brera, Istituto Nazionale di Astrofisica (INAF), 23807 Merate, Italy.}
\affil[2]{Dipartimento di Scienza e Alta Tecnologia (DiSAT), Universit\`a degli studi dell'Insubria, 22100 Como, Italy.}
\affil[3]{Mets\"{a}hovi Radio Observatory, Aalto University, 02540 Kylm\"{a}l\"{a}, Finland.}
\affil[4]{Th\"uringer Landessternwarte, D-07778 Tautenburg, Germany;
on leave of absence from: Departamento de Astronom\'ia, Universidad de Guanajuato, Guanajuato 36023, GTO, Mexico.}
\affil[5]{Osservatorio Astronomico di Padova, Istituto Nazionale di Astrofisica (INAF), 35122 Padova, Italy.}
\affil[6]{Department of Astronomy, Boston University, Boston, MA 02215 USA.}
\affil[7]{Homer L. Dodge Department of Physics and Astronomy, University of Oklahoma, Norman, OK 73019 USA.}
\affil[8]{CFisUC, Departamento de F\'{i}sica, Universidade de Coimbra, 3004-516 Coimbra, Portugal.}
\affil[9]{Dipartimento di Fisica e Astronomia, Universit\`a di Padova, 35122 Padova, Italy.}

\begin{document}
\maketitle

\abstract{We present the results of a comparison between different methods to estimate the power of relativistic jets from active galactic nuclei (AGN). We selected a sample of 32 objects (21 flat-spectrum radio quasars, 7 BL Lacertae Objects, 2 misaligned AGN, and 2 changing-look AGN) from the Very Large Baseline Array (VLBA) observations at 43~GHz of the Boston University blazar program. We then calculated the total, radiative, and kinetic jet power from both radio and high-energy gamma-ray observations, and compare the values. We found an excellent agreement between the radiative power calculated by using the Blandford \& K\"onigl model with 37 or 43~GHz data, and the values derived from the high-energy $\gamma-$ray luminosity. The agreement is still acceptable if 15~GHz data are used, although with a larger dispersion, but it improves if we use a constant fraction of the $\gamma-$ray luminosity. We found a good agreement also for the kinetic power calculated with Blandford \& K\"onigl model with 15~GHz data, and the value from the extended radio emission. We also propose some easy-to-use equations to estimate the jet power. \textbf{Keywords:} Relativistic Jets; Active Galactic Nuclei; Seyfert Galaxies; BL Lac Objects; Flat-Spectrum Radio Quasars.}

\section{Introduction}
Accreting compact objects can emit powerful relativistic jets (see \cite{BLANDFORD2019,FOSCHINI2022A} for recent reviews on jets from active galactic nuclei, AGN). One key quantity to understand the physics of jets, and its impact on the nearby environment (either the host galaxy and/or the intergalactic medium), is the power -- both radiative and kinetic -- that is dissipated in these structures. There are many ways to estimate the power based on different observational quantities, but the results are generally not consistent, with differences of one or more orders of magnitude. Despite this clear mismatch between the various methods, very few works have been published to understand the origin of this problem. 

Pjanka et al. \cite{PJANKA2017} compared four methods: a one-zone leptonic model by Ghisellini et al. \cite{GHISELLINI2009,GHISELLINI2014}, radio core shifts \cite{LOBANOV1998,ZDZIARSKI2015}, extended radio emission (radio lobes or steep radio spectrum) \cite{WILLOT1999}, and the high-energy gamma-ray luminosity \cite{NEMMEN2012,GHISELLINI2014}. They found that the powers estimated according to the leptonic model and the radio core shifts are almost consistent, while the value derived from the $\gamma-$ray luminosity is about half, and that from the radio lobes is about one order of magnitude smaller. These calculations can be reconciled by taking into account the source variability across time (the power derived from extended radio emission is an average over the source lifetime) or a change in the ratio between the number of leptons to hadrons (at least 15-to-1), or in the magnetization of the jet (and giving up the ideal magnetohydrodynamic theory). However, Pjanka et al. concluded that they are unable to decide which option is best. 

We too made a preliminary study by comparing the interpretations of the same method by different authors \cite{FOSCHINI2019}. Therefore, we compared the radiative power derived from the $\gamma-$ray luminosity, and the Lorentz and Doppler factors from radio observations at different frequencies from \cite{JORSTAD2017,PUSHKAREV2017,LIODAKIS2018,FINKE2019}. We compared the broad-band spectral modeling by \cite{GHISELLINI2014,PALIYA2017,PALIYA2019}, and the observation of extended radio emission by \cite{MEYER2011,NOKHRINA2015}. We also compared the relationships by \cite{FOSCHINI2014}, based on the 15 GHz radio luminosity, with radiative power from the $\gamma-$ray luminosity. Although the models by Ghisellini \cite{GHISELLINI2014} and Paliya \cite{PALIYA2017,PALIYA2019} were described as the same (one-zone leptonic model), their results are systematically different toward low power values, with Paliya's values being about one order of magnitude greater than Ghisellini's. Something similar was found also by comparing the power calculated from the extended radio emission, with differences of one to two orders of magnitudes at low powers. In this case, the reason was likely due to the different approaches: while Nokhrina \cite{NOKHRINA2015} directly considered observations at 326~MHz, Meyer \cite{MEYER2011} started from 1.4~GHz observations and extrapolated to 300~MHz. The latter is not suitable to estimate the steep-spectrum radio emission from lobes, because it fades as the frequency increases, and extended emission might not be detected already at GHz. The comparison of Foschini's relationship between jet power and 15~GHz radio core luminosity (see Eq.~1 in \cite{FOSCHINI2014}), with the radiative power from $\gamma$ rays plus Lorentz and Doppler factors from radio observations at 43 GHz by Jorstad \cite{JORSTAD2017}, resulted in a fair agreement, although with significant dispersion. 

One major limitation of our previous work was to compare published works. Therefore, we could not select the epochs of observations, change models, or reanalyse data. In the present work, we overcome these limitations, and address, in some more detail, the estimation of the jet power from a small, but reliable sample of jetted AGN. Our aim is to understand the reasons of discrepancies, and, if possible, to propose solutions. We also search for easy-to-use solutions, which might be of great value for the analysis of large samples of objects. A simple relationship between the power and an observed quantity, or an equation linking a few observed quantities is easier to use than a detailed, but complex, numerical model. Obviously, the discrepancies have to be smaller than one order of magnitude to be acceptable.

We adopted the most recent value of the Hubble constant for the local Universe, $H_0=73.3$~km~s$^{-1}$~Mpc$^{-1}$ from \cite{RIESS2022}, and calculated the luminosity distance $d_{\rm L}$ by using the simplified equation:

\begin{equation}
d_{\rm L} \sim \frac{cz}{H_0}(1+\frac{z}{2}) \,\, \mathrm{[Mpc]}
\label{distance}
\end{equation}

\noindent where $c$ is the speed of light in vacuum, and $z$ is the redshift.

Since we are comparing different methods based on the same data, we did not consider measurement errors, which are often quite large, but we focused on the dispersion $\sigma$ of the values.

\begin{table}[ht!] 
\caption{\footnotesize Sample of jetted AGN derived from \cite{JORSTAD2017}. Column explanation: (1) IAU source name referred to J2000; (2) a more common alias; (3) Right Ascension ([deg], J2000); (4) Declination ([deg], J2000); (5) classification (BLLAC: BL Lac Object; MIS: misaligned AGN; FSRQ: flat-spectrum radio quasar; CLAGN: changing-look AGN); (6) redshift. Information for columns (5) and (6) were taken from \cite{FOSCHINI2022B}. \label{sourcelist}}
\centering
\vskip 6pt
\begin{tabular}{lcrrcc}
\hline
\textbf{Name}	& \textbf{Alias}	& \textbf{RA} & \textbf{Dec} & \textbf{Class} & \textbf{z}\\
(1) & (2) & (3) & (4) & (5) & (6)\\
\hline
J$0238+1636$		& PKS~$0235+164$	& $39.66$ & $+16.62$	& BLLAC 	& $0.940$\\
J$0319+4130$		& NGC~$1275$		& $49.95$ & $+41.51$	& MIS 		& $0.0176$\\
J$0339-0146$        & PKS~$0336-01$ 	& $54.88$ & $-1.78$	& FSRQ 		& $0.852$\\
J$0423-0120$        & PKS~$0420-01$		& $65.82$ & $-1.34$ 	& FSRQ 		& $0.915$\\
J$0433+0521$        & 3C~$120$ 			& $68.30$ & $+5.35$ 	& MIS 		& $0.0336$ \\
J$0530+1331$        & PKS~$0528+134$ 	& $82.73$ & $+13.53$ 	& FSRQ 		& $2.07$\\
J$0830+2410$        & S3~$0827+24$      & $127.72$ & $+24.18$    & FSRQ      & $0.941$\\
J$0831+0429$        & PKS~$0829+046$    & $127.95$ & $+4.49$    & BLLAC     & $0.174$\\
J$0841+7053$        & 4C~$+71.07$       & $130.35$ & $+70.89$ 	& FSRQ      & $2.17$\\
J$0854+2006$        & OJ~$287$          & $133.70$ & $+20.11$ 	& BLLAC     & $0.306$\\
J$0958+6533$        & S4~$0954+65$      & $149.70$ & $+65.56$    & BLLAC     & $0.368$\\
J$1058+0133$        & 4C~$+01.28$       & $164.62$ & $+1.57$    & FSRQ      & $0.892$\\
J$1104+3812$        & Mkn~$421$         & $166.11$ & $+38.21$    & BLLAC     & $0.0308$\\
J$1130-1449$        & PKS~$1127-145$    & $172.53$ & $-14.82$	& FSRQ      & $1.19$\\
J$1159+2915$        & Ton~$599$	 		& $179.88$ & $+29.24$    & FSRQ      & $0.725$\\
J$1221+2813$        & W~Comae           & $185.38$ & $+28.23$	& BLLAC     & $0.102$\\
J$1224+2122$        & 4C~$+21.35$       & $186.23$ & $+21.38$ 	& FSRQ      & $0.434$\\
J$1229+0203$        & 3C~$273$          & $187.28$ & $+2.05$ 	& FSRQ      & $0.158$\\
J$1256-0547$        & 3C~$279$          & $194.05$ & $-5.79$    & FSRQ      & $0.536$\\
J$1310+3220$        & OP~$313$      	& $197.62$ & $+32.34$	& FSRQ      & $0.996$\\
J$1408-0752$        & PKS~B$1406-076$   & $212.24$ & $-7.87$ 	& FSRQ      & $1.49$\\
J$1512-0905$        & PKS~$1510-089$    & $228.21$ & $-9.10$ 	& FSRQ      & $0.360$\\
J$1613+3412$        & OS~$319$          & $243.42$ & $+34.21$	& FSRQ      & $1.40$\\
J$1626-2951$        & PKS~B$1622-297$   & $246.52$ & $-29.86$ 	& FSRQ      & $0.815$\\
J$1635+3808$        & 4C~$+38.41$       & $248.81$ & $+38.13$ 	& FSRQ      & $1.81$\\
J$1642+3948$        & 3C~$345$          & $250.74$ & $+39.81$	& FSRQ      & $0.593$\\
J$1733-1304$        & PKS~$1730-13$     & $263.26$ & $-13.08$   	& FSRQ      & $0.902$\\
J$1751+0939$        & OT~$081$          & $267.89$ & $+9.65$ 	& CLAGN     & $0.320$\\
J$2202+4216$        & BL~Lac            & $330.68$ & $+42.28$	& BLLAC     & $0.0686$\\
J$2225-0457$        & 3C~$446$          & $336.45$ & $-4.95$  	& CLAGN     & $1.40$\\
J$2232+1143$        & CTA~$102$         & $338.15$ & $+11.73$ 	& FSRQ      & $1.04$\\
J$2253+1608$        & 3C~$454.3$        & $343.49$ & $+16.15$	& FSRQ      & $0.858$\\
\hline
\end{tabular}
\end{table}

\section{Sample Selection}
\label{sample}
We selected the sample of the Very Long Baseline Array (VLBA) Boston University (BU) blazar program\footnote{Now BEAM-ME, \url{https://www.bu.edu/blazars/BEAM-ME.html}} \cite{BUBLAZAR,JORSTAD2017}. It is composed of 36 objects observed with VLBA at 43 GHz between 2007 June and 2013 January. We cross matched this sample with the catalogue of revised classifications and redshifts for the jetted AGN sample in the fourth {\em Fermi} Large Area Telescope (LAT) catalog (4FGL) as published by \cite{FOSCHINI2022B}. We thus removed four objects: 3C~66A, S5~$0716+71$, and PKS~$0735+17$, because they have no spectroscopic redshift (only estimates from photometry or the imaging of the host galaxy), and 3C~111, because its Galactic latitude is $|b|\leq 10^{\circ}$ and therefore not included in the above cited work.

The remaining 32 objects are listed in Table~\ref{sourcelist}, and were classified in \cite{FOSCHINI2022B} as follows: 21 flat-spectrum radio quasars (FSRQ), 7 BL Lac Objects (BLLAC), 2 misaligned AGN (MIS, also known as radio galaxies), and 2 changing-look AGN (CLAGN). The latter type has different meanings, depending on the authors. It was originally introduced by Matt et al. \cite{MATT2003} to indicate AGN switching from Compton-thin to Compton-thick obscuration. In more recent years, also changes in the accretion were considered (e.g. \cite{RICCI2022}). In the present case, CLAGN indicates jetted AGN with optical spectra showing dramatic changes, from a featureless continuum to a line-dominate spectrum, or vice versa, thus moving from one class to another (for example, from BLLAC to FSRQ, and/or vice versa; cf. \cite{FOSCHINI2022B}). We kept in the sample both MIS and CLAGN to avoid reducing too much a small sample and to have some insight on how large viewing angles and dramatic changes in the electromagnetic emission can affect the jet power.

It is worth noting that there are some slight differences in the values of the redshift with respect to \cite{JORSTAD2017}. Therefore, we recalculated the affected quantities (e.g. the brightness temperature) to take into account these changes. This is mostly for the sake of consistency, rather than a significant change of the affected quantities.

In addition to the Boston University blazar program, there is also another excellent VLBA program: the Monitoring Of Jets in Active galactic nuclei with VLBA Experiments (MOJAVE\footnote{\url{https://www.cv.nrao.edu/MOJAVE/}}, \cite{LISTER2018}). We cross-matched the above cited sample with the larger sample (447 AGN) of the MOJAVE program \cite{HOMAN2021}, which offers a comparable set of physical quantities measured from radio observations at 15~GHz or derived from them. All 32 objects from the BU blazar program have been observed in the MOJAVE program. Also in this case, we found some cases of slightly different redshift, and we corrected the affected quantities. 

\section{The Blandford \& K\"onigl model}
The first step is to use the simplified and evergreen model by Blandford \& K\"onigl \cite{BLANDFORD1979} to estimate the jet power. For the sake of simplicity, we shortly recall the main concepts and refer to the above cited work \cite{BLANDFORD1979} for more details. Blandford \& K\"onigl considered a conical jet, with opening semiangle $\phi$, and the axis inclined by an angle $\theta$ with respect to the line of sight to the observer, so that the observed opening angle is $\phi_{\rm obs}=\phi/ \sin \theta$. The jet is a stream of relativistic electrons with distribution: 

\begin{equation}
N(\gamma_{\rm e}) = K \gamma_{\rm e}^{-2}
\label{electrondist}
\end{equation}

\noindent where $K$ is a normalization constant, and $\gamma_{\rm e}$ is the random Lorentz factor of the electrons in the range $\gamma_{\rm e,min}<\gamma_{\rm e}<\gamma_{\rm e,max}$. The magnetic field $B$ is tangled with the plasma, and the bulk motion of the electrons has constant speed $\beta$ (in units of $c$), linked to the measured apparent speed $\beta_{\rm app}$ via:

\begin{equation}
\beta = \frac{\beta_{\rm app}}{\beta_{\rm app}\cos \theta + \sin \theta}
\label{betarel}
\end{equation}

The electron energy density is:

\begin{equation}
u_{\rm e} = K m_{\rm e} c^2 \log (\frac{\gamma_{\rm e,max}}{\gamma_{\rm e,min}})
\label{electronenergy}
\end{equation}

\noindent where $m_{\rm e}$ is the electron rest mass, while the energy density of the magnetic field is:

\begin{equation}
u_{\rm B} = \frac{B^2}{8\pi}
\label{magneticenergy}
\end{equation}

\noindent Equipartition between $u_{\rm e}$ and $u_{\rm B}$ is assumed, via the constant $k_{\rm eq}$, generally smaller than 1 (Blandford \& K\"onigl assumed $k_{\rm eq}=0.5$ in their example \cite{BLANDFORD1979}).

Blandford \& K\"onigl then calculated the expected flux density at radio frequencies, given the power of the jet (Eq.~29 in \cite{BLANDFORD1979}):

\begin{equation}
S_{\nu}\sim \frac{1}{2}(1+z)k_{\rm eq}^{\frac{5}{6}}\Delta^{-\frac{17}{12}}(1+\frac{2}{3}k_{\rm eq}\Lambda)^{-\frac{17}{12}}\Gamma^{-\frac{17}{6}}\beta^{-\frac{17}{12}}\delta^{\frac{13}{6}}(\sin \theta)^{-\frac{5}{6}} \phi_{\rm obs}^{-1}P_{\rm 44}^{\frac{17}{12}}d_{\rm L,9}^{-2} \, \, [\rm Jy]
\label{fluxobs}
\end{equation}
 
\noindent where $S_{\nu}$ is the observed flux density at the frequency $\nu$, $\Delta=\log (r_{\rm max}/r_{\rm min})$, where $r_{\rm min}$ and $r_{\rm max}$ refer to the size of the emission region, $\Lambda=\log (\gamma_{\rm e,max}/\gamma_{\rm e,min})$, $\Gamma$ is the bulk Lorentz factor, $\delta$ is the Doppler factor, $d_{\rm L,9}$ is the luminosity distance in units of Gpc, and $P_{44}$ is the total jet power in units of $10^{44}$~erg~s$^{-1}$. We can rearrange the Eq.~(\ref{fluxobs}) to calculate the jet power as a function of the observed radio flux density and the other observed physical quantities:

\begin{equation}
P_{\rm 44}\sim k_{1}k_{2} \left(\frac{S_{\nu}d_{\rm L,9}^2}{1+z}\right)^{12/17}
\label{totaljetpow}
\end{equation}

\noindent where the factor $k_{1}$ depends on the electron random Lorentz factors and the size of the emitting region:

\begin{equation}
k_1 = \left(\frac{1}{2}\right)^{-12/17} k_{\rm eq}^{-10/17}\Delta (1+\frac{2}{3}k_{\rm eq}\Lambda) 
\label{fudge1}
\end{equation}

\noindent while $k_{2}$ depends on the observed quantities:

\begin{equation}
k_{2}= \Gamma^2 \beta \delta^{-26/17} (\sin \theta)^{10/17} \phi_{\rm obs}^{12/17}
\label{fudge2}
\end{equation}

The synchrotron radiative power is:

\begin{equation}
P_{\rm rad,syn,44}\sim \frac{k_{\rm eq}}{2(1+\frac{2}{3}k_{\rm eq}\Lambda)}P_{\rm 44}
\label{synchropow}
\end{equation}

By adopting the typical values suggested by Blandford \& K\"onigl \cite{BLANDFORD1979} for $k_{\rm eq}=0.5$, $\Delta = 5$, and $\Lambda = 3$, we obtain $k_{1}\sim 24.5$. Therefore:

\begin{equation}
P_{\rm 44}\sim 24.5 k_{2} \left(\frac{S_{\nu}d_{\rm L,9}^2}{1+z}\right)^{12/17}
\label{totaljetpow}
\end{equation}

\begin{equation}
P_{\rm rad,syn,44}\sim \frac{1}{8} P_{\rm 44}
\label{synchropow}
\end{equation}

It immediately follows that the jet kinetic power is:

\begin{equation}
P_{\rm kin,44}\sim \frac{7}{8} P_{\rm 44}
\label{kinpow}
\end{equation}

\section{Very Long Baseline Array (VLBA) observations}
\label{VLBA}

\subsection{All epochs}
\label{VLBA1}
The data collected by the BU blazar program span from June 2007 to January 2013, while the MOJAVE program covers the years from 1994 to 2019. The first check was based on all the data available (Table~\ref{allepochs}). We noted that one object in the MOJAVE sample (J$0238+1636$) has no measurement of $\beta_{\rm app}$: therefore, we adopted the value from Jorstad et al. \cite{JORSTAD2017}. The MOJAVE program has also no measurement of the $\phi_{\rm obs}$, and therefore we calculated it by means of the relationship $\Gamma\phi \sim 0.1-0.2$ \cite{CLAUSEN2013,PJANKA2017}. We adopted $\Gamma\phi \sim 0.11$ as suggested by \cite{PJANKA2017}, but tested also the case of $\Gamma\phi \sim 0.2$, resulting in no significant changes. Since we need the observed opening angle $\phi_{\rm obs}$ in Eq.~(\ref{fluxobs}), the above cited relationship can be rewritten as:

\begin{equation}
\phi_{\rm obs} = \frac{0.11}{\Gamma \sin \theta}
\label{openingangle}
\end{equation}

This equation was used also to calculate $\phi_{\rm obs}$ at 43~GHz for J$0238+1636$, because this measurement was missing.

We distinguished two cases. In {\em case 1}, the Doppler factor at 43~GHz was derived from the flux density variability of the jet knots, according to Eq.~(3) in \cite{JORSTAD2017}:

\begin{equation}
\delta = \frac{15.8 s d_{\rm L,9}}{\tau(1+z)}
\label{dopplerBU}
\end{equation}

\noindent where $s$ is the angular size of the knot (mas), and $\tau$ is the variability time scale (years). Therefore, we corrected Eq.~(\ref{dopplerBU}) to take into account the slightly different values of $z$ and $d_{\rm L,9}$. 

\begin{table}[h!]
\centering
\begin{threeparttable}
\caption{\footnotesize Input data corrected for different redshifts and $H_0$ (all epochs). Columns description: (1) source name (J2000); (2) median flux density at 15~GHz [Jy]; (3) 15~GHz brightness temperature [K]; (4) maximum apparent speed as measured from 15~GHz observations [$c$]; (5) median flux density at 43~GHz [Jy]; (6) 43~GHz brightness temperature [K]; (7) maximum apparent speed as measured from 43~GHz observations [$c$]; (8) Doppler factor as measured according to Eq.~(\ref{dopplerBU}); (9) observed jet opening semiangle [deg]. Original data at 15 and 43~GHz are taken from \cite{HOMAN2021} and \cite{JORSTAD2017}, respectively. To avoid reducing too much the small sample, we considered the few cases of lower limits as detections.} 
\label{allepochs}
\begin{tabular}{lcccccccc}
\hline
Name & $S_{\rm 15\, GHz}$ & $\log T_{\rm b,15}$ & $\beta_{\rm max,15}$ & $S_{\rm 43\, GHz}$ & $\log T_{\rm b,43}$ & $\beta_{\rm max,43}$ & $\delta_{43}$ & $\phi_{\rm obs}$ \\
(1) & (2) & (3) & (4) & (5) & (6) & (7) & (8) & (9) \\
\hline
J$0238+1636$ & 1.33 & 11.70 & 26.27\tnote{1} & 1.77 & 10.97\tnote{2} & 26.27 & 48.75 & 5.84\tnote{3} \\
J$0319+4130$ & 2.84 & 11.25 & 0.41 & 15.51 & 10.79 & 0.36 & 10.37 & 22.1 \\
J$0339-0146$ & 1.56 & 12.03 & 24.5 & 1.68 & 11.19\tnote{2} & 31.42 & 14.39 & 7.2 \\
J$0423-0120$ & 4.80 & 12.49 & 5.46 & 4.74 & 11.90 & 15.53 & 15.56 & 23.4 \\
J$0433+0521$ & 0.958 & 11.36 & 6.28 & 1.67 & 11.46 & 8.7 & 4.33 & 6.6 \\
J$0530+1331$ & 2.18 & 12.14 & 18.41 & 2.02 & 11.91 & 77.94 & 20.79 & 22.8 \\
J$0830+2410$ & 1.19 & 11.78 & 19.8 & 1.28 & 11.47 & 18.04 & 21.03 & 24.0 \\
J$0831+0429$ & 0.525 & 11.39 & 10.2 & 0.57 & 10.97 & 7.23 & 12.33 & 7.9 \\
J$0841+7053$ & 1.50 & 12.14 & 21.51 & 1.73 & 11.20 & 25.15 & 16.80 & 6.8 \\
J$0854+2006$ & 2.74 & 12.27 & 15.14 & 4.68 & 11.88 & 8.6 & 7.9 & 33.0 \\
J$0958+6533$ & 0.903 & 11.76 & 14.8 & 1.05 & 11.45 & 17.58 & 7.78 & 21.0 \\
J$1058+0133$ & 3.57 & 12.50 & 6.61 & 4.02 & 11.61 & 14.14 & 18.42 & 24.8 \\
J$1104+3812$ & 0.319 & 11.14 & 0.218 & 0.28 & 10.18 & 1.07 & 23.42 & 55.2 \\
J$1130-1449$ & 1.12 & 11.80 & 19.8 & 1.76 & 11.36 & 23.37 & 19.68 & 15.4 \\
J$1159+2914$ & 1.57 & 11.95 & 24.6 & 1.40 & 11.59 & 15.47 & 10.75 & 13.6 \\
J$1221+2813$ & 0.226 & 11.31 & 8.2 & 0.25 & 10.79 & 4.76 & 8.96 & 9.2 \\
J$1224+2122$ & 1.40 & 11.83 & 21.8 & 1.19 & 11.66 & 13.81 & 6.72 & 16.2 \\
J$1229+0203$ & 3.52 & 11.95 & 14.91 & 11.88 & 12.51 & 11.83 & 3.97 & 6.6 \\
J$1256-0547$ & 11.94 & 12.76 & 20.5 & 18.05 & 11.92 & 16.01 & 16.54 & 47.4 \\
J$1310+3220$ & 1.55 & 11.99 & 27.5 & 2.14 & 10.95 & 13.73 & 19.37 & 58.4 \\
J$1408-0752$ & 0.802 & 12.06 & 22.77 & 0.59 & 11.32 & 29.42 & 10.89 & 16.4 \\
J$1512-0905$ & 1.87 & 11.95 & 28.0 & 2.44 & 11.15 & 29.6 & 31.98 & 11.4 \\
J$1613+3412$ & 2.73 & 12.25 & 31.1 & 1.53 & 11.09 & 9.82 & 7.29 & 20.8 \\
J$1626-2951$ & 0.959 & 12.01 & 12.0 & 1.35 & 11.34 & 11.04 & 8.68 & 30.8 \\
J$1635+3808$ & 2.02 & 12.46 & 30.8 & 2.93 & 11.86 & 10.17 & 12.71 & 41.2 \\
J$1642+3948$ & 3.27 & 12.29 & 19.37 & 4.47 & 11.63 & 19.45 & 10.77 & 18.6 \\
J$1733-1304$ & 3.07 & 12.29 & 27.3 & 3.31 & 11.92 & 23.52 & 7.27 & 16.2 \\
J$1751+0939$ & 3.52 & 12.62 & 6.85 & 3.60 & 11.65 & 17.66 & 15.97 & 26.2 \\
J$2202+4216$ & 2.28 & 11.87 & 10.0 & 4.21 & 11.99\tnote{2} & 11.89 & 7.00 & 6.0 \\
J$2225-0457$ & 4.75 & 12.30 & 17.7 & 3.82 & 11.62 & 22.20 & 13.19 & 22.0 \\
J$2232+1143$ & 2.04 & 12.38 & 20.0 & 2.71 & 11.59 & 27.93 & 28.49 & 23.8 \\
J$2253+1608$ & 3.53 & 12.26 & 17.0 & 14.44 & 12.22 & 9.06 & 22.35 & 22.6 \\
\hline
\end{tabular}
\begin{tablenotes}
\footnotesize
\item[1] Missing. Value from 43 GHz measurements.
\item[2] Lower limit.
\item[3] Missing. Value from Eq.~(\ref{openingangle}).
\normalsize
\end{tablenotes}
\end{threeparttable}
\end{table}

Then, $\Gamma$ and $\theta$ were also updated according to the well-known equations (e.g. \cite{JORSTAD2017,HOMAN2021}):

\begin{equation}
\Gamma = \frac{\beta_{\rm max}^2+\delta^2+1}{2\beta_{\rm max}}
\label{lorentzf}
\end{equation}

\begin{equation}
\theta = \arctan \frac{2\beta_{\rm max}}{\beta_{\rm max}^2+\delta^2-1} 
\label{viewingangle}
\end{equation}

In {\em case 2}, we tested the effect of recalculating the Doppler factor at 43~GHz by using the brightness temperature ratio:

\begin{equation}
\delta = \frac{T_{\rm b, 43}}{T_{\rm b,int}}
\label{dopplertb}
\end{equation}

\noindent where $T_{\rm b, 43}$ is the observed brightness temperature [K] (see Table~\ref{allepochs}), and $T_{\rm b,int}=5\times 10^{10}$~K is the theoretical intrinsic value \cite{READHEAD}. We underline that case 1 and case 2 differ in the calculation of $\delta$ at 43~GHz (Eq.~\ref{dopplerBU} vs Eq.~\ref{dopplertb}). The Doppler factor at 15~GHz is always derived from the brightness temperature. It is also worth noting that we adopt $\beta_{\rm max}$ as the reference apparent speed, because it correlates better with $T_{\rm b}$, as suggested by \cite{HOMAN2021}. 

\begin{figure}[h]
\centering
\includegraphics[width=7 cm]{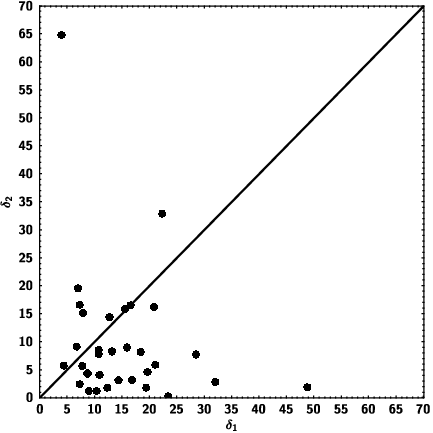}
\caption{\footnotesize Doppler factor $\delta_1$ (case 1) estimated from 43~GHz data and Eq.~(\ref{dopplerBU}) with the cosmology adopted in the present work vs $\delta_2$ (case 2) calculated from the brightness temperature. The continuous line indicates the equality of the two values.\label{dopplerBU2}}
\end{figure} 

Fig.~\ref{dopplerBU2} shows the comparison of $\delta$ as measured with the two cited methods. We noted some cases with extreme differences: J$0238+1636$, $\delta_1\sim 49$, $\delta_2\sim 2$; J$1104+3812$, $\delta_1\sim 23$, $\delta_2\sim 0.30$; J$1229+0203$, $\delta_1\sim 4$, $\delta_2\sim 65$. The reasons might be that, e.g., J$0238+1636$ has no measured $\phi_{\rm obs}$, and its $T_{\rm b,43}$ is a lower limit, and J$1229+0203$ has the highest $T_{\rm b,43}$. This might imply the breakdown of the equipartition assumption for the observed brightness temperature $T_{\rm b,obs}>10^{13}$~K, as already noted by \cite{READHEAD,JORSTAD2017}. The case of J$1104+3812$ might be due to the so-called Doppler crisis in BL Lacs \cite{EDWARDS2002,PINER2004}. 

\begin{figure}[h]
\centering
\includegraphics[width=6.5 cm]{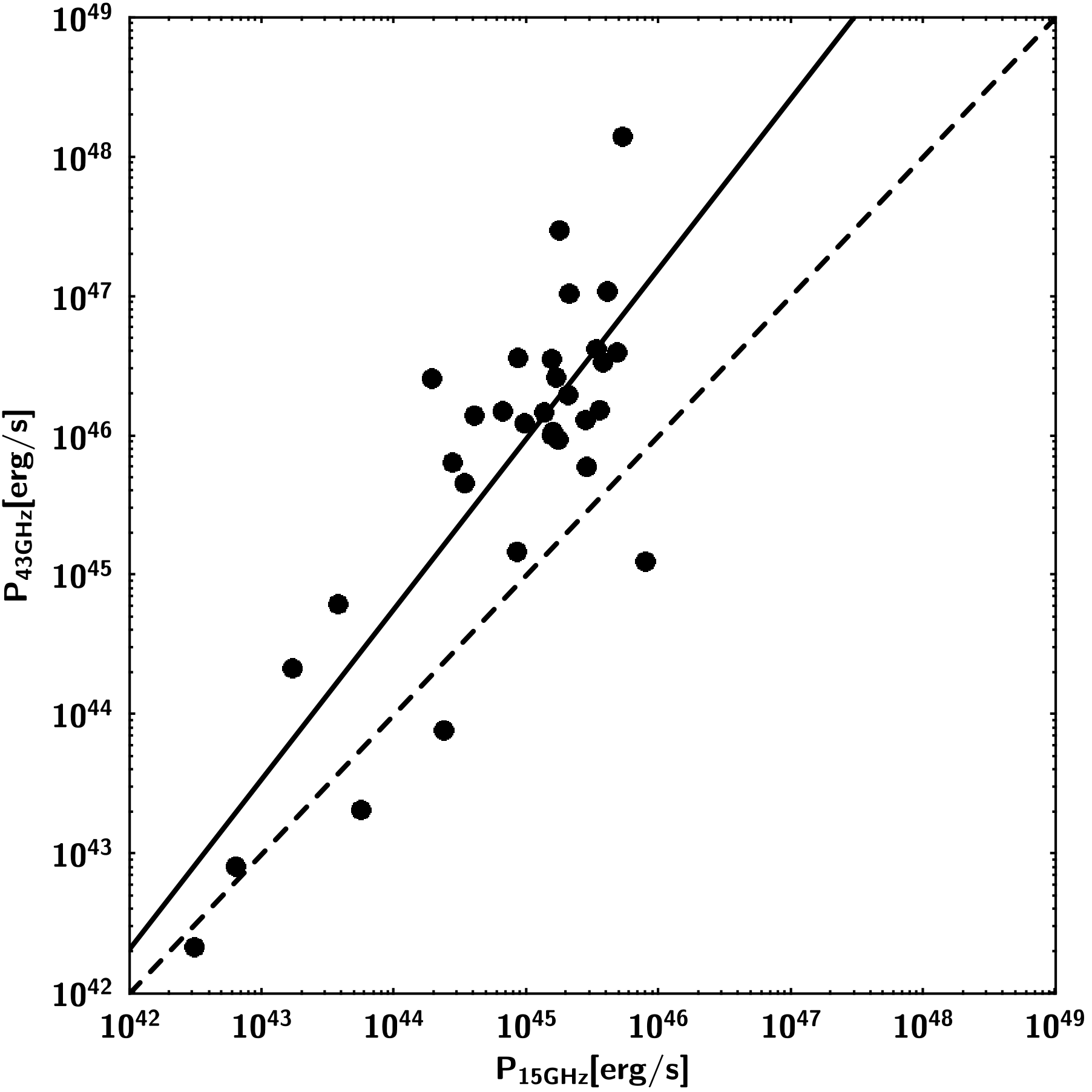}
\includegraphics[width=6.5 cm]{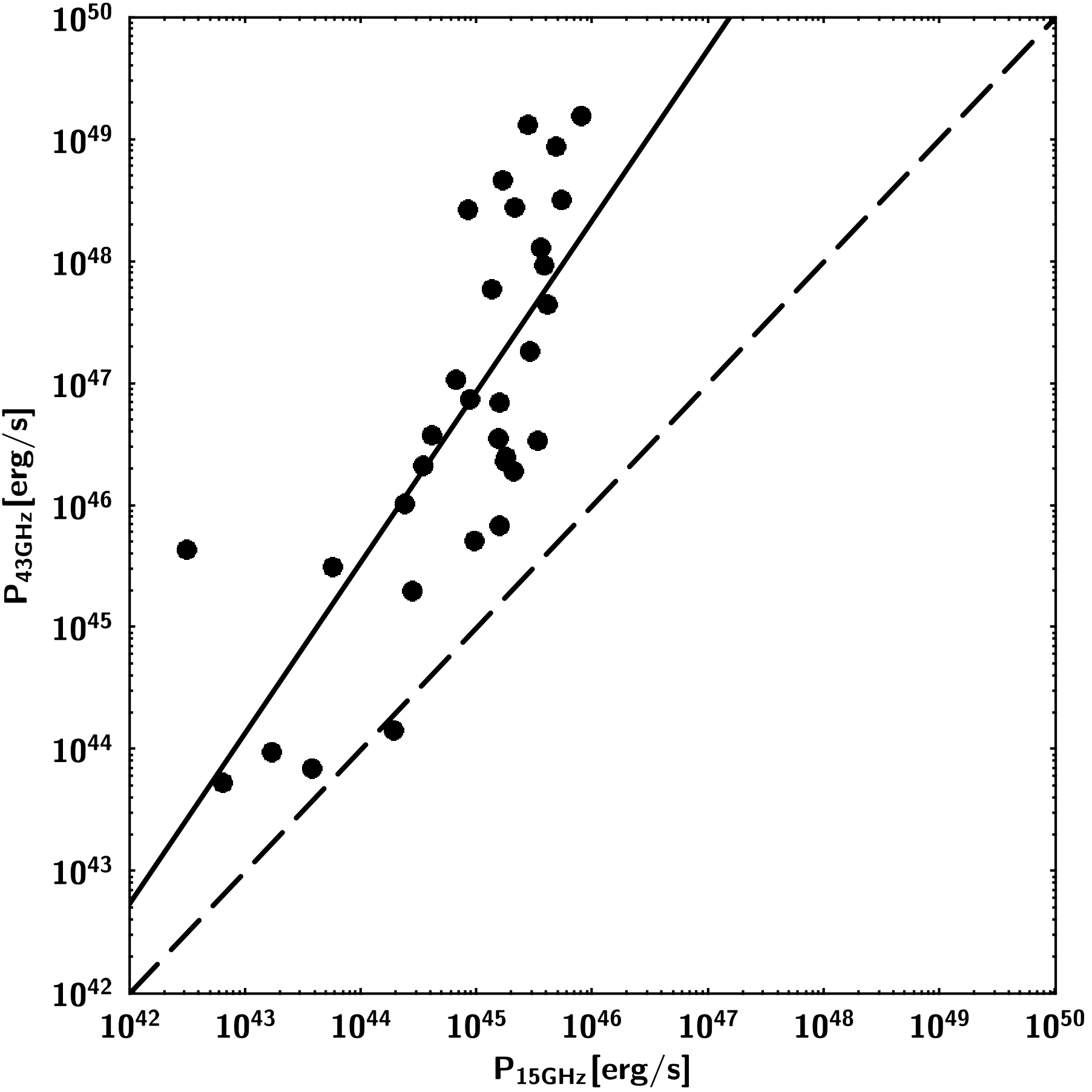}
\caption{\footnotesize Total jet power calculated from Eq.~(\ref{totaljetpow}) and all the data from 15 and 43~GHz observations. ({\it left panel}) case 1;  ({\it right panel}) case 2. The dashed line represents the equality of the two powers, while the continuous line is the linear fit to the data.\label{radioall}}
\end{figure}   

\begin{figure}[h]
\centering
\includegraphics[width=7 cm]{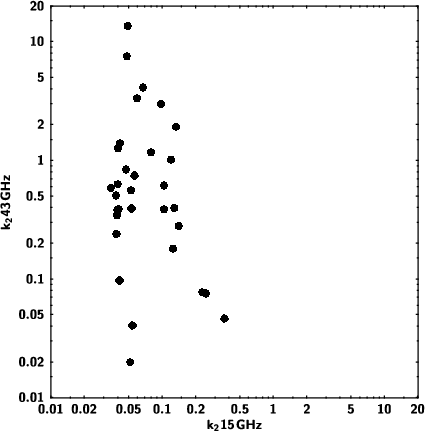}
\caption{\footnotesize Distribution of $k_{2}$ values at 15 and 43~GHz for the case 1. \label{kappa2}}
\end{figure}

Fig.~\ref{radioall} displays the total jet power in the two cases and compared with the values derived from 15~GHz data. It is also worth studying the distribution of the coefficients $k_{2}$ (see Eq.~\ref{fudge2}), which is shown in Fig.~\ref{kappa2} for case 1. 

The mean value of $k_{2}$ is $0.088$ ($\sigma\sim 0.072$), and $1.4$ ($\sigma\sim 2.6$) for 15 and 43~GHz, respectively (please note that $k_{2}>0$ by definition, so the dispersion is mostly toward values greater than the average). In the case 2 (not shown), the dispersion increases to $\sim 131$, while the mean value rises to $\sim 61$. It is worth noting that the distribution of $k_{2}$ for 15~GHz data is quite narrow, with all the values between $\sim 0.034$ and $\sim 0.36$.

It is evident that the derivation of the Doppler factor from the brightness temperature at 43~GHz (case 2) leads to a more pronounced divergence at higher powers and a larger dispersion. The linear fit in the form:

\begin{equation}
\log P_{\rm 43\, GHz} = m\log P_{\rm 15\, GHz} + C
\label{linfit}
\end{equation}

\noindent gives the following values: $m\sim 1.22$, and $C\sim -8.9$ for case 1; $m\sim 1.40$, and $C\sim -16$ for case 2. The correlation factor $\rho$ is $0.82$, and $0.77$, for case 1 and case 2, respectively, while the dispersion $\sigma$ is $0.72$, and $0.96$. 

Nonetheless, the relatively small range of $k_2$ values (particularly at 15~GHz, see Fig.~\ref{kappa2}) offers an interesting possibility to derive the jet power only on the basis of the flux density at radio frequencies, although some caveats must be taken into account (see Sect.~\ref{FF}).

\subsection{Overlapping epochs}
\label{VLBA2}
\begin{figure}[h]
\centering
\includegraphics[width=6.5 cm]{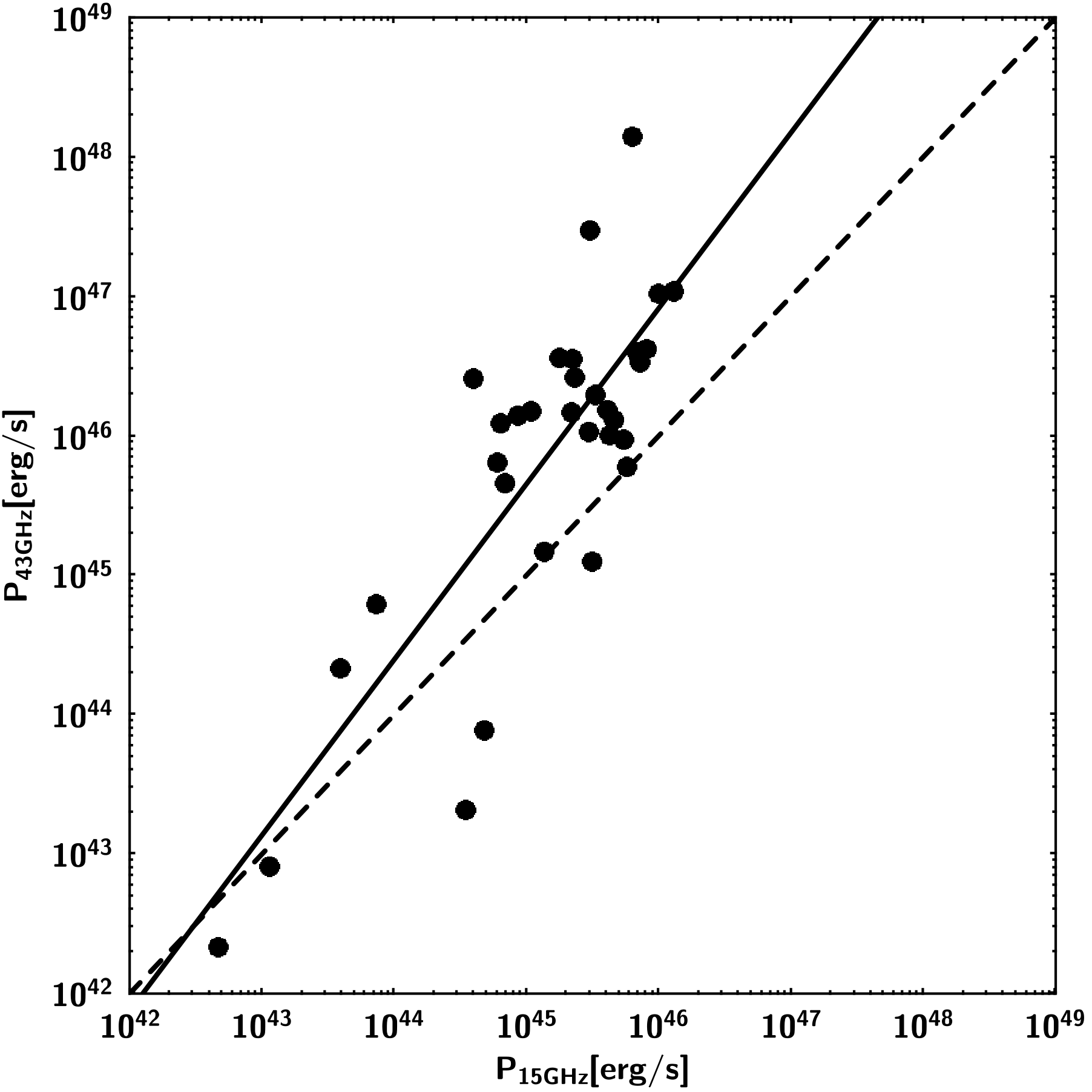}
\includegraphics[width=6.5 cm]{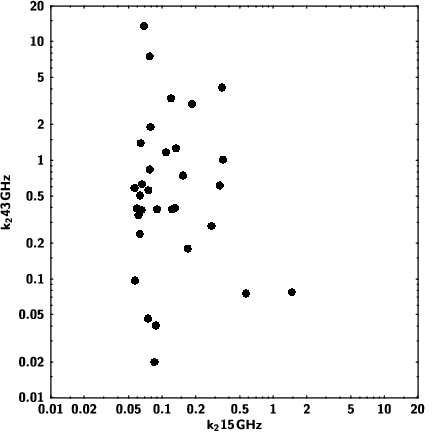}
\caption{\footnotesize ({\it left panel}) Total jet power calculated with Eq.~(\ref{totaljetpow}) and overlapping epochs data from 15 and 43~GHz observations. The dashed line line represents the equality of the two powers, while the continuous line is the linear fit to the data.  ({\it right panel}) Distribution of $k_{2}$ values at 15 and 43~GHz. \label{radiooverlap}}
\end{figure}

We remind the reader that 43~GHz data span from June 2007 to January 2013 \cite{JORSTAD2017}, while the 15~GHz data cover the years from 1994 to 2019 \cite{HOMAN2021}. In the previous subsection, we considered all the available epochs, but now we want to study the case of overlapping epochs. Therefore, we collected 15~GHz data only if observed between June 1, 2007, and January 31, 2013 (Table~\ref{overlapdata}). The results are shown in Fig.~\ref{radiooverlap}.

\begin{table}[h!]
\caption{\footnotesize Input data corrected for different redshifts and $H_0$ (overlapping epochs, from June 2007 to January 2013). Columns description: (1) source name (J2000); (2) median flux density at 15~GHz [Jy]; (3) 15~GHz brightness temperature [K]; (4) median flux density at 37~GHz [Jy]. Original data at 15~GHz from \cite{HOMAN2021}. See Sect.~\ref{metsahovi} for 37~GHz data of the Mets\"ahovi Radio Observatory. \label{overlapdata}}
\centering
\vskip 6pt
\begin{tabular}{lccccc}
\hline
Name & $S_{\rm 15\, GHz}$ & $\log T_{\rm b,15}$ & $S_{\rm 37\, GHz}$ \\
(1) & (2) & (3) & (4) \\
\hline
J$0238+1636$ & 3.37 & 12.31 & 1.50 \\
J$0319+4130$ & 3.26 & 11.27 & 17.34 \\
J$0339-0146$ & 1.58 & 12.09 & 2.33 \\
J$0423-0120$ & 4.65 & 12.27 & 5.18  \\
J$0433+0521$ & 0.675 & 11.22 & 1.95 \\
J$0530+1331$ & 1.69 & 12.26 & 1.69 \\
J$0830+2410$ & 1.23 & 11.74 & 1.42 \\
J$0831+0429$ & 0.434 & 11.32 & 0.724 \\
J$0841+7053$ & 2.10 & 12.56 & 2.20 \\
J$0854+2006$ & 3.99 & 12.27 & 5.01 \\
J$0958+6533$ & 1.07 & 11.72 & 1.19 \\
J$1058+0133$ & 4.37 & 12.56 & 4.25 \\
J$1104+3812$ & 0.292 & 11.12 & 0.428 \\
J$1130-1449$ & 1.32 & 11.94 &  $-$\\
J$1159+2914$ & 1.53 & 11.79 & 1.61 \\
J$1221+2813$ & 0.216 & 11.06 & 0.363 \\
J$1224+2122$ & 1.68 & 12.21 & 1.69 \\
J$1229+0203$ & 3.66 & 11.86 & 16.49 \\
J$1256-0547$ & 9.82 & 12.72 & 18.67 \\
J$1310+3220$ & 2.44 & 12.09 & 2.20 \\
J$1408-0752$ & 0.708 & 11.75 & 0.812 \\
J$1512-0905$ & 2.38 & 12.00 & 2.62 \\
J$1613+3412$ & 1.61 & 12.07 & 2.35 \\
J$1626-2951$ & 0.959 & 12.02 & $-$ \\
J$1635+3808$ & 2.27 & 12.26 & 3.62 \\
J$1642+3948$ & 4.86 & 12.37 & 5.69 \\
J$1733-1304$ & 3.32 & 12.30 & 3.69 \\
J$1751+0939$ & 4.70 & 12.72 & 3.38 \\
J$2202+4216$ & 3.44 & 11.98 & 4.50 \\
J$2225-0457$ & 4.56 & 11.89 & 3.43 \\
J$2232+1143$ & 2.11 & 12.46 & 2.79 \\
J$2253+1608$ & 9.16 & 12.76 & 7.21 \\
\hline
\end{tabular}
\end{table}

There are no significant changes with respect to the previous cases. The linear fit gives these parameters: $m\sim 1.26$, $C\sim -11$, $\rho\sim 0.82$, $\sigma\sim 0.71$. However, we note an increase of the mean value of $k_2$ at 15~GHz and its dispersion, from $\sim 0.088$ ($\sigma\sim 0.072$) to $\sim 0.18$ ($\sigma\sim 0.26$). Since $k_2$ is a function of $\beta$, $\Gamma$, $\delta$, $\theta$, and $\phi_{\rm obs}$ (see Eq.~\ref{fudge2}), a change in the observing epochs results in a different median $T_{\rm b,15\,GHz}$, which in turn affects $\delta$, and all the other parameters of $k_{2}$. The possibility of having a greater or smaller mean value and dispersion depends on the activity of the objects during the selected time interval. Anyway, in the present case, the distribution is still narrow, with only two values greater than the previous limit of $\sim 0.36$. The two objects are J$0831+0429$ ($k_2\sim 0.57$), and J$1221+2813$ ($k_2\sim 1.5$). Given the lack of significant changes with respect to $P_{\rm 43\, GHz}$, we concluded that changes in $k_{2}$ were partially compensated by changes in flux density.

\section{Single-dish observations}
\label{metsahovi}
The next test is to use the above calculated $k_{2}$ factors to estimate the jet power from single-dish observations. This type of observations does not allow to measure or to derive all the quantities necessary to calculate $k_{2}$ (which are $\delta,\Gamma,\beta,\theta,\phi_{\rm obs}$, cf Eq.~\ref{fudge2}): it is possible to measure only $\delta$ from the brightness temperature \cite{HOVATTA2009,LIODAKIS2018}, but then it is necessary to take $\beta_{\rm app}$ from VLBA observations to derive the other quantities according to the Eqs.~(\ref{openingangle}), (\ref{lorentzf}), and (\ref{viewingangle}). Therefore, we can try to use $k_2$ as measured from the above cited VLBA observations coupled to the flux density as measured from single-dish observations. 

Data from the Mets\"ahovi Radio Observatory\footnote{\url{https://www.metsahovi.fi/opendata/}} (MRO) of the Aalto University (Finland) were used. MRO is a $\sim 14$~m single dish, equipped with a 1~GHz-band dual-beam receiver centered at $36.8$~GHz. The high electron mobility pseudomorphic transistor (HEMPT) front end operates at ambient temperature. The observations, with typical exposures of $\sim 10^3$~s, are Dicke switched ON--ON observations, alternating between the source and the sky in each feed horn. The detection threshold is $\sim 0.2$~Jy, in the best case. Calibration sources were the HII region DR 21, NGC~7027, 3C~274, and 3C~84. More information about data reduction and analysis can be found in \cite{MROREF}.

All the objects in Table~\ref{sourcelist} were monitored for more than 30~years, with the exception of J$1130-1449$ and J$1626-2951$. For the sake of simplicity, we considered only the case of overlapping epochs (see Table~\ref{overlapdata}). 

\begin{figure}[h]
\centering
\includegraphics[width=6.5 cm]{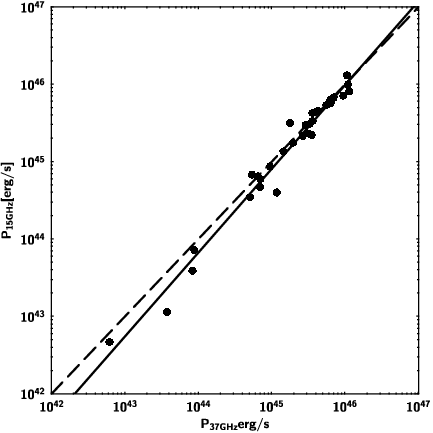}
\includegraphics[width=6.5 cm]{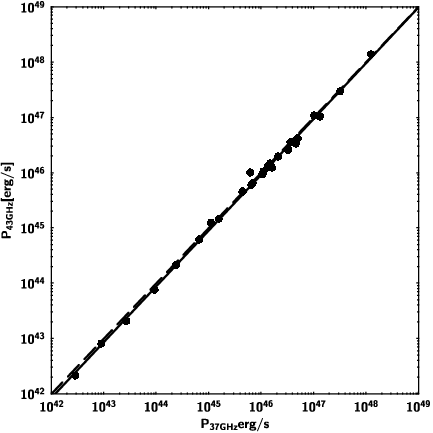}
\caption{\footnotesize ({\it left panel}) Total jet power derived from 37~GHz flux density and $k_2$ from 15~GHz observations vs jet power from the same observations. ({\it right panel}) Total jet power calculated by using 37~GHz flux density and $k_2$ from 43~GHz observations vs jet power from the same observations.  The dashed line represents the equality of the two powers, while the continuous line is the linear fit to the data. \label{mro}}
\end{figure}  

The results are displayed in Fig.~\ref{mro}. We note a very good correlation between the new values of jet power from MRO at 37~GHz and the values of MOJAVE (15~GHz) and BU (43~GHz), with a best result if $k_2$ is measured from VLBA observations at the closer frequency (43~GHz), as expected. The linear fit (cf Eq.~\ref{linfit}) gives the following results: 

\begin{itemize}
\item	$k_2$ from MOJAVE (15~GHz): $m\sim 1.08$, $C\sim -3.8$, $\rho\sim 0.99$, $\sigma\sim 0.14$;
\item	$k_2$ from BU (43~GHz): $m\sim 1.00$, $C\sim -0.46$, $\rho\sim 1.00$, $\sigma\sim 0.067$;
\end{itemize}

\section{Kinetic power estimated from the extended emission}
\label{kinetic}
The extended radio emission offers the opportunity to estimate the kinetic power of the jet. McNamara et al. \cite{MCNAMARA2000} found a deficit of X-ray emission from the surrounding cluster at the location of the radio lobes of Hydra A, indicating that the jet had excavated cavities in the intergalactic medium. Then, by studying these X-ray cavities of a sample of radio galaxies in clusters, B\^{i}rzan et al. \cite{BIRZAN2008} found a correlation between the jet kinetic power and its extended radio emission at $327$~MHz:

\begin{equation}
\log \frac{P_{\rm kin}}{10^{42}}=0.51\log \frac{P_{\rm 327\,MHz}}{10^{40}}+1.51
\label{birzan}
\end{equation}

\noindent where $P_{\rm kin}$ is the jet kinetic power [erg/s], while $P_{\rm 327\,MHz}$ is the radio power as measured at $327$~MHz [erg/s]. Later, Cavagnolo et al. \cite{CAVAGNOLO2010} enlarged the sample by adding also isolated giant elliptical galaxies, and proposed a new relationship based on the extended radio emission measured at $200-400$~MHz:

\begin{equation}
\log \frac{P_{\rm kin}}{10^{42}}=0.64\log \frac{P_{\rm 200-400\,MHz}}{10^{40}}+1.54
\label{cavagnolo}
\end{equation}

\noindent where $P_{\rm 200-400\,MHz}$ is the radio power as measured at $200-400$~MHz [erg/s]. The authors also proposed a relationship with the radio power as measured at $1.4$~GHz, but this is less reliable \cite{BIRZAN2008,CAVAGNOLO2010,FOSCHINI2019}, and therefore we do not consider it. 

\begin{table}[h!]
\centering
\begin{threeparttable}
\caption{\footnotesize Input data for the kinetic power. Columns description: (1) source name (J2000); (2) median flux density at $327$~MHz [Jy]; (3) median flux density at $200-400$~MHz [Jy]. All the data were extracted from the CATS database \cite{CATS}.}
\label{fluxmhz}
\begin{tabular}{lcc}
\hline
Name & $S_{\rm 327\, MHz}$ & $S_{\rm 200-400\, MHz}$ \\
(1) & (2) & (3) \\
\hline
J$0238+1636$ & 1.04 & 1.26\\
J$0319+4130$ & 42.8\tnote{1} & 27.06\\
J$0339-0146$ & 0.943 & 1.33\\
J$0423-0120$ & 0.820 & 1.20\\
J$0433+0521$ & 2.37 & 6.33\\
J$0530+1331$ & 1.13\tnote{1} & 1.05\\
J$0830+2410$ & 0.660\tnote{2} & 0.770\\
J$0831+0429$ & 1.19\tnote{2} & 0.837\\
J$0841+7053$ & 5.07 & 5.07\\
J$0854+2006$ & 0.790 & 1.15\\
J$0958+6533$ & 0.624\tnote{1} & 0.742\\
J$1058+0133$ & 4.39 & 4.42\\
J$1104+3812$ & 0.961 & 1.14\\
J$1130-1449$ & 4.51\tnote{3} & 5.35\\
J$1159+2914$ & 3.52 & 2.71\\
J$1221+2813$ & 1.45 & 0.790\\
J$1224+2122$ & 3.98\tnote{2} & 4.80\\
J$1229+0203$ & 62.89 & 64.0\\
J$1256-0547$ & 14.79 & 14.58\\
J$1310+3220$ & 1.43 & 1.42\\
J$1408-0752$ & 0.535\tnote{4} & 0.584\\
J$1512-0905$ & 2.51 & 2.73\\
J$1613+3412$ & 2.55 & 3.11\\
J$1626-2951$ & 2.37\tnote{4} & 2.46\\
J$1635+3808$ & 2.51 & 2.31\\
J$1642+3948$ & 9.93 & 8.70\\
J$1733-1304$ & 4.66 & 7.61\\
J$1751+0939$ & 1.17\tnote{2} & 0.720\\
J$2202+4216$ & 1.82\tnote{1} & 2.77\\
J$2225-0457$ & 12.71 & 12.15\\
J$2232+1143$ & 6.99 & 7.88\\
J$2253+1608$ & 11.67 & 12.44\\
\hline
\end{tabular}
\begin{tablenotes}
\footnotesize 
\item[1] From 325~MHz observations.
\item[2] From 318~MHz observations.
\item[3] From 333~MHz observations.
\item[4] From 227~MHz observations.
\normalsize
\end{tablenotes}
\end{threeparttable}
\end{table}

To measure the radio power we followed the procedure outlined in \cite{CAVAGNOLO2010}, and extracted the radio data from the CATS database\footnote{\url{https://www.sao.ru/cats/}} \cite{CATS}. As noted by Cavagnolo \cite{CAVAGNOLO2010}, it is difficult to find $327$~MHz data for all the objects, and therefore the search was extended to the range $200-400$~MHz. In the case of our sample, we found 327~MHz data for 21/32 objects. To avoid reducing too much our small sample, we used radio data at close frequencies ($227,318,325,333$~MHz) when $327$~MHz data were not available. In the case of $200-400$~MHz, we also considered the cited frequency range with a tolerance of $\pm10$\%. We performed the K-correction of the radio fluxes by adopting an average spectral index $\alpha=0.8$ ($S_{\nu}\propto \nu^{-\alpha}$), as done by \cite{CAVAGNOLO2010}. We did not restrict the selected data from observations in the period $2007-2013$, because the time necessary to excavate cavities in the intergalactic medium is of the order of several $10^8$ years (e.g. \cite{MCNAMARA2000}). Therefore, the measure of the kinetic power refers to an average over a very long time scale. The flux densities are displayed in Table~\ref{fluxmhz}. 

Fig.~\ref{kinetic} shows the comparisons of the jet kinetic power as calculated with Eqs.~(\ref{birzan}) and (\ref{cavagnolo}). The two values are well correlated ($\rho\sim 0.99$, $\sigma\sim 0.084$), but there is an evident divergence at high radio powers ($m\sim 1.24$, $C\sim -10.6$). This is somehow expected, given the different slopes of the two relationships ($0.64/0.51\sim 1.25$, cf Eqs.~\ref{birzan} and \ref{cavagnolo}). The reason for this divergence might be the different samples adopted by B\^irzan \cite{BIRZAN2008} and Cavagnolo \cite{CAVAGNOLO2010}: while the former built the correlation by selecting a sample of radio galaxies in clusters (where, given the density and temperature of the intergalactic gas, it is easier to detect X-ray cavities), the latter added also a group of isolated giant elliptical galaxies (where X-ray cavities might be more difficult to detect). 

\begin{figure}[h]
\centering
\includegraphics[width=7 cm]{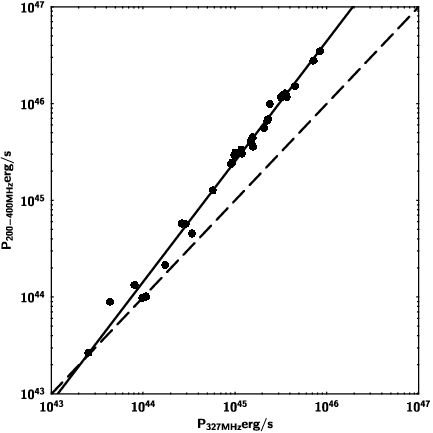}
\caption{\footnotesize Comparison of the jet kinetic power as estimated from Eqs.~(\ref{birzan}) and (\ref{cavagnolo}). The dashed line represents the equality of the two powers, while the continuous line is the linear fit to the data. \label{kinetic}}
\end{figure}  

\begin{figure}[h]
\centering
\includegraphics[width=6.5 cm]{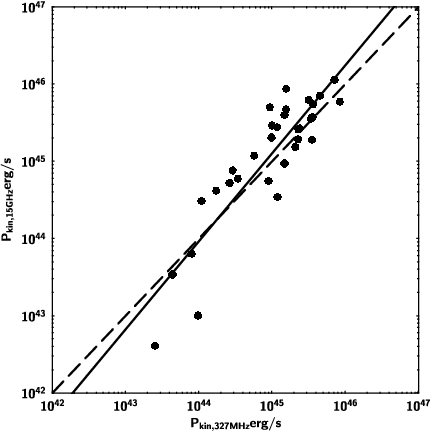}
\includegraphics[width=6.5 cm]{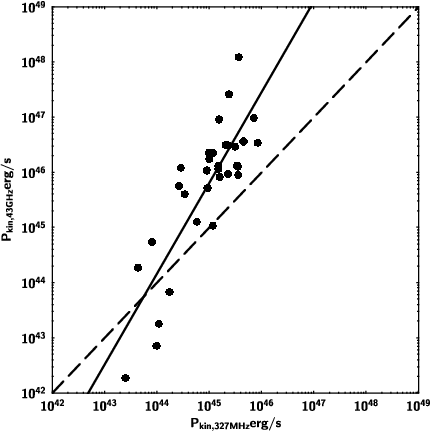}\\
\includegraphics[width=6.5 cm]{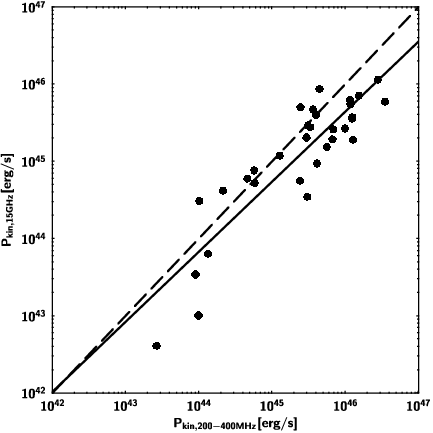}
\includegraphics[width=6.5 cm]{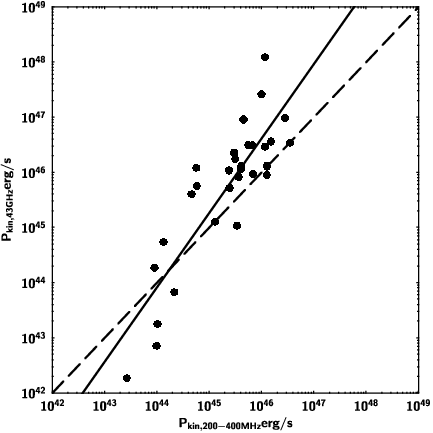}
\caption{\footnotesize Kinetic jet power. ({\it upper panels}) Comparison of Eq.~(\ref{birzan}) and  Eq.~(\ref{kinpow}) with 15~GHz data ({\it left}), and 43~GHz data ({\it right}). ({\it lower panels}) Comparison of Eq.~(\ref{cavagnolo}) and Eq.~(\ref{kinpow}) with 15~GHz data ({\it left}), and 43~GHz data ({\it right}). The dashed line represents the equality of the two powers, while the continuous line is the linear fit to the data. \label{radioext}}
\end{figure}

Fig.~\ref{radioext} displays the four comparisons between the kinetic power calculated with Eq.~(\ref{kinpow}) and data from 15 or 43~GHz observations, and the values calculated with Eq.~(\ref{birzan}) or Eq.~(\ref{cavagnolo}) with the measurements of the extended radio emission at MHz frequencies. 

The linear fits give these values:

\begin{itemize}
\item	327~MHz vs 15~GHz: $m\sim 1.14$, $C\sim -6.2$, $\rho\sim 0.89$, $\sigma\sim 0.37$;
\item	327~MHz vs 43~GHz: $m\sim 1.65$, $C\sim -28$, $\rho\sim 0.84$, $\sigma\sim 0.68$;
\item	200-400~MHz vs 15~GHz: $m\sim 0.91$, $C\sim 3.9$, $\rho\sim 0.89$, $\sigma\sim 0.38$;
\item	200-400~MHz vs 43~GHz: $m\sim 1.35$, $C\sim -15.5$, $\rho\sim 0.86$, $\sigma\sim 0.64$;
\end{itemize}

All the powers are well correlated ($\rho\sim 0.84-0.89$), showing a smaller dispersion when using 15~GHz data.  In all cases, we noted a systematic underestimation of the power as calculated with Eq.~(\ref{kinpow}) for weak sources, with $P_{\rm kin}\lesssim 10^{44}$~erg/s (or an overestimation of the relationships based on the extended radio emission). The comparison with 43~GHz data shows a clear divergence toward higher radio powers. One source of bias is the fact that we used the integrated flux density, while we should have taken only the steep spectrum emission of the lobes. However, since our sources have moderately to high redshift (with a few exceptions), it is not possible to disentangle the core from the lobes. 

We would like also to note that Eqs.~(\ref{birzan}) and (\ref{cavagnolo}), are not the result of a theoretical calculation, but are correlations derived from observed quantities. Therefore, as is well known that correlation is not causation, the above cited relationships heavily rely on the adopted samples, as also shown by the change in the slope from Eq.~(\ref{birzan}) to Eq.~(\ref{cavagnolo}) displayed in Fig.~\ref{kinetic}. 

\section{Radiative power}
\label{LAT}
The last test is with the radiative power as measured at high-energy $\gamma$ rays by the {\em Fermi} Large Area Telescope (LAT) \cite{ATWOOD2009}. Since all the versions of the LAT catalogs cover a time span greater than the Boston University program \cite{JORSTAD2017}, we extracted the data from the {\em Fermi} LAT Light Curve Repository\footnote{\url{https://fermi.gsfc.nasa.gov/ssc/data/access/lat/LightCurveRepository/index.html}} \cite{LATLCR} covering only the epoch of the Boston University program ($2007-2013$). This web site is an automatic generator of light curves based on the likelihood with a power-law model, and with a limited selection of parameters. We selected one month time bin and left the photon index free to vary. We extracted the light curves starting from the beginning of LAT operations (2008 August) until 2013 January, and then calculated the weighted mean of the observed $0.1-100$~GeV flux $F_{\gamma}$ and the spectral index $\alpha_{\gamma}$ (Table~\ref{gammadata}).

\begin{table}[h!]
\caption{\footnotesize {\em Fermi}/LAT data in the period August 2008 - January 2013. Columns description: (1) source name (J2000); (2) $0.1-100$~GeV flux [$10^{-11}$~erg~cm$^{-2}$~s$^{-1}$]; (3) spectral index $\alpha_{\gamma}$. All the data were downloaded from the {\em Fermi} LAT Light Curve Repository \cite{LATLCR}.\label{gammadata}}
\centering
\vskip 6pt
\begin{tabular}{lcc}
\hline
Name & $F_{\rm 0.1-100\, GeV}$ & $\alpha_{\gamma}$ \\
(1) & (2) & (3) \\
\hline
J$0238+1636$ & 10.0 & 1.20\\
J$0319+4130$ & 22.0 & 1.07\\
J$0339-0146$ & 4.6 & 1.50\\
J$0423-0120$ & 5.5 & 1.40\\
J$0433+0521$ & 1.5 & 1.70\\
J$0530+1331$ & 3.5 & 1.60\\
J$0830+2410$ & 3.2 & 1.70\\
J$0831+0429$ & 4.0 & 1.20\\
J$0841+7053$ & 3.4 & 1.80\\
J$0854+2006$ & 6.4 & 1.20\\
J$0958+6533$ & 1.6 & 1.40\\
J$1058+0133$ & 8.2 & 1.20\\
J$1104+3812$ & 44.0 & 0.73\\
J$1130-1449$ & 2.1 & 1.60\\
J$1159+2914$ & 8.2 & 1.30\\
J$1221+2813$ & 4.0 & 1.20\\
J$1224+2122$ & 30.0 & 1.60\\
J$1229+0203$ & 18.0 & 2.00\\
J$1256-0547$ & 23.0 & 1.40\\
J$1310+3220$ & 2.8 & 1.50\\
J$1408-0752$ & 2.1 & 1.40\\
J$1512-0905$ & 52.0 & 1.46\\
J$1613+3412$ & 1.3 & 1.40\\
J$1626-2951$ & 2.7 & 1.70\\
J$1635+3808$ & 20.0 & 1.40\\
J$1642+3948$ & 4.6 & 1.20\\
J$1733-1304$ & 6.0 & 1.50\\
J$1751+0939$ & 4.4 & 1.30\\
J$2202+4216$ & 17.0 & 1.28\\
J$2225-0457$ & 2.1 & 1.60\\
J$2232+1143$ & 14.0 & 1.50\\
J$2253+1608$ & 174 & 1.50\\
\hline
\end{tabular}
\end{table}

From these values, we calculated the $0.1-100$~GeV luminosity:

\begin{equation}
L_{\gamma} = 4\pi d_{\rm L}^2\frac{F_{\gamma}}{(1+z)^{1-\alpha_{\gamma}}}
\label{gamma}
\end{equation}

The minimum radiative power $P_{\mathrm{rad},\gamma}$ from high-energy $\gamma$ rays (i.e. via inverse-Compton scattering) can be estimated as follows \cite{MARASCHI2003}:

\begin{equation}
P_{\mathrm{rad},\gamma} \sim \frac{\Gamma^2}{\delta^4}L_{\gamma}
\label{radgamma}
\end{equation}

\begin{figure}[h]
\centering
\includegraphics[width=6.5 cm]{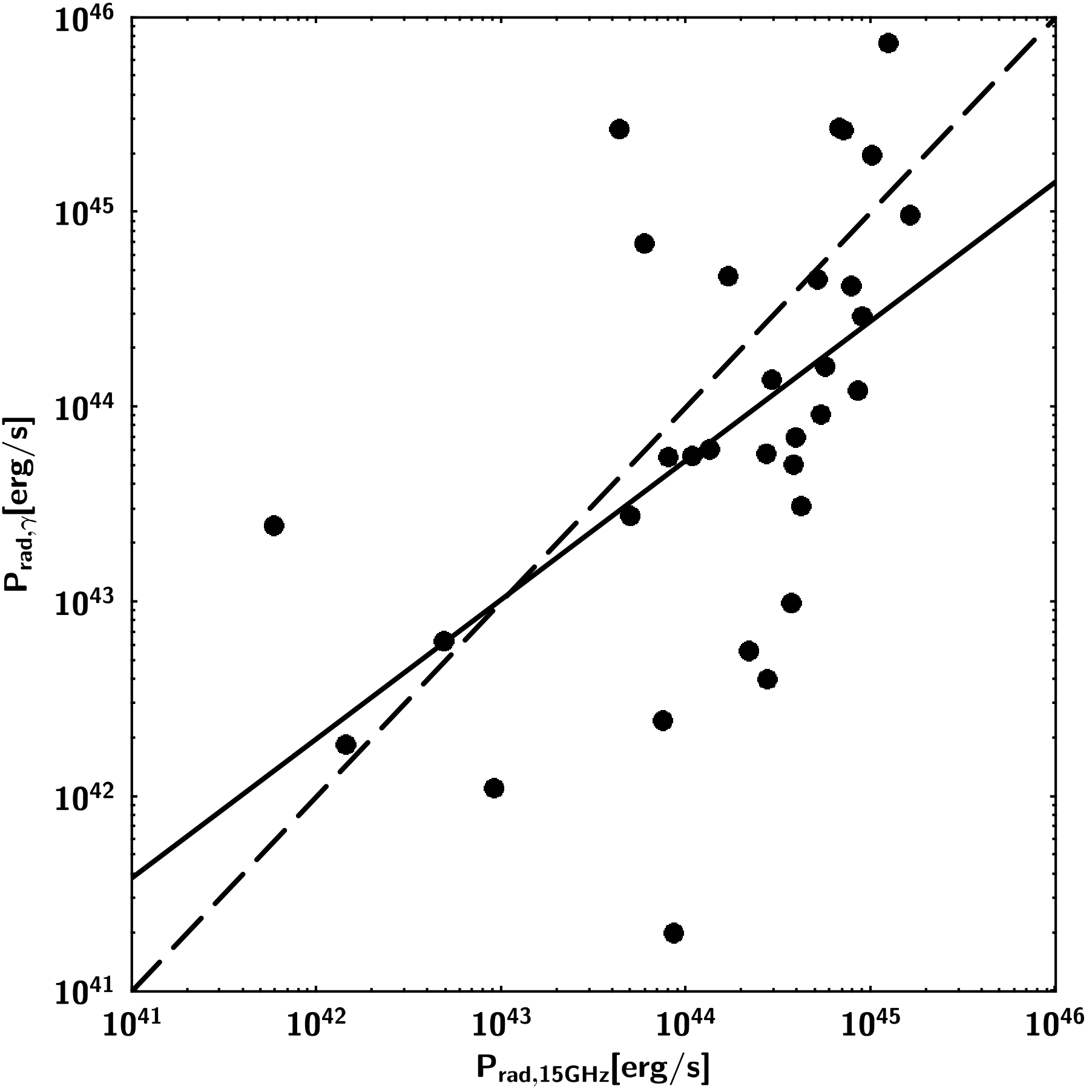}
\includegraphics[width=6.5 cm]{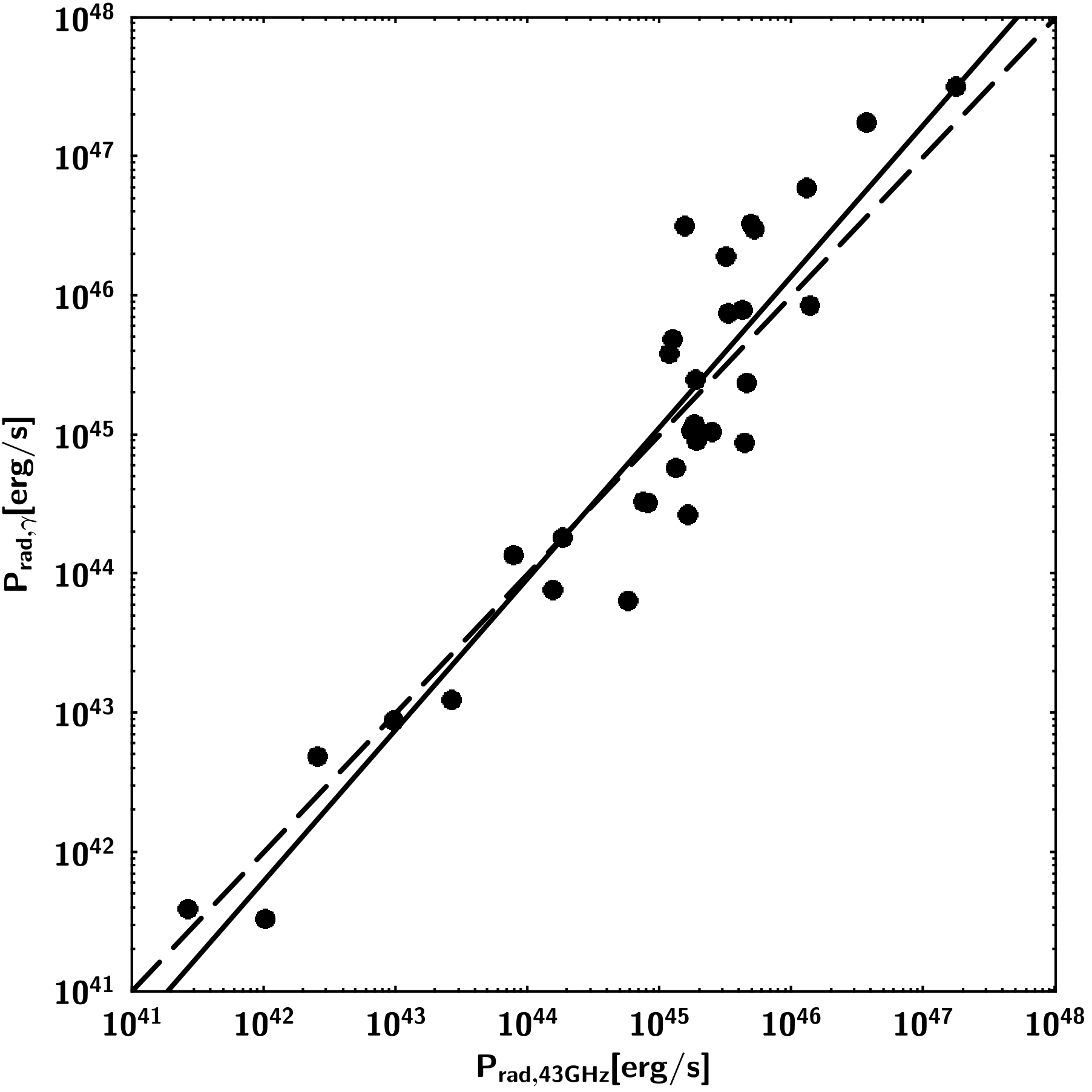}
\caption{\footnotesize Radiative jet power. Comparison of values from high-energy $\gamma$ rays and radio observations at 15~GHz ({\it left panel}), and at 43~GHz ({\it right panel}). The dashed line represents the equality of the two powers, while the continuous line is the linear fit to the data.  \label{radiativep}}
\end{figure}

The values of $\Gamma$ and $\delta$ can be taken from the VLBA observations at 15 and 43~GHz. Then, $P_{\mathrm{rad},\gamma}$ can be compared with the synchrotron radiative power calculated according to Eq.~(\ref{synchropow}). The results are displayed in Fig.~\ref{radiativep}.

We note a good agreement, with a smaller dispersion when using 43~GHz data. The results of the linear fits are:

\begin{itemize}
\item	$\gamma$ rays vs 15~GHz: $m\sim 0.71$, $C\sim 12$, $\rho\sim 0.54$, $\sigma\sim 0.92$;
\item	$\gamma$ rays vs 43~GHz: $m\sim 1.09$, $C\sim -3.9$, $\rho\sim 0.94$, $\sigma\sim 0.51$;
\end{itemize}

It is worth noting that Eq.~(\ref{synchropow}) calculates the radiative power emitted via the synchrotron process, while the radiative power measured at high-energy $\gamma$ rays can have a significant contribution from the external Compton process in FSRQs. It is known (e.g. \cite{BOTTCHER}) that the total power radiated by relativistic electrons is:

\begin{equation}
P_{\rm rad,tot}=P_{\rm rad,syn}+P_{\mathrm{rad},\gamma} = \frac{4}{3}\sigma_{\rm Th}c\gamma_{e}^2 u_{\rm B}(1+k_{CD})
\end{equation}

\noindent where $P_{\rm rad,syn}$ is the power dissipated via synchrotron radiation, $P_{\mathrm{rad},\gamma}$ is the power due to the inverse-Compton process, and $\sigma_{\rm Th}\sim 0.66\times 10^{-28}$~m$^{2}$ is the Thompson cross section. The Compton dominance parameter $k_{\rm CD}$ is defined as follows:

\begin{equation}
k_{\rm CD}= \frac{u_{\rm seed}}{u_{\rm B}}
\label{comptdom}
\end{equation}

\noindent where $u_{\rm seed}$ is the energy density of the seed photons field (from accretion disk, broad-line region, molecular torus, ...). The Compton dominance can be measured from the observations of the peaks of synchrotron and inverse-Compton emissions:

\begin{equation}
k_{\rm CD}\sim \frac{\nu F_{\nu}^{\rm IC}}{\nu F_{\nu}^{\rm syn}}
\label{comptdomobs}
\end{equation}

\begin{figure}[h]
\centering
\includegraphics[width=6.5 cm]{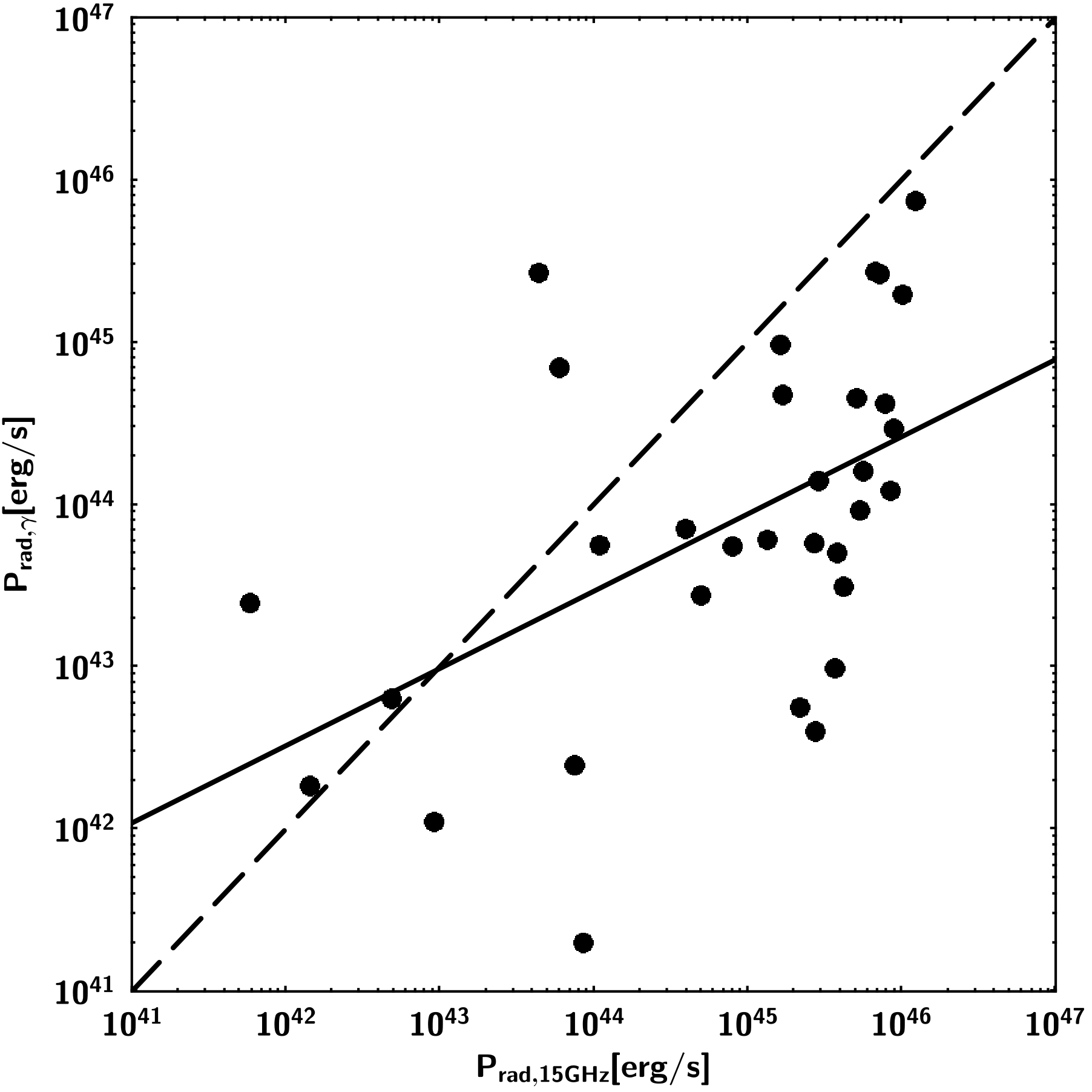}
\includegraphics[width=6.5 cm]{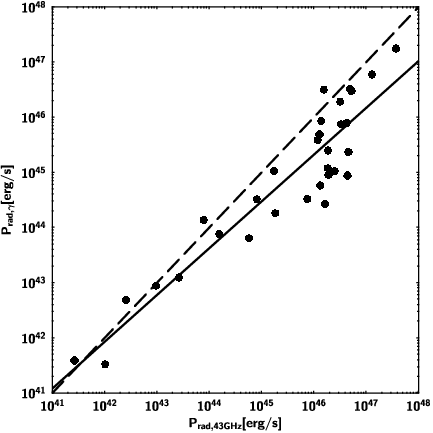}
\caption{\footnotesize Radiative jet power corrected for the Compton dominance. Comparison of values from high-energy $\gamma$ rays and radio observations at 15~GHz ({\it left panel}), and at 43~GHz ({\it right panel}). The dashed line represents the equality of the two powers, while the continuous line is the linear fit to the data.  \label{cdradpow}}
\end{figure}

From the inspection of a large sample of spectral energy distributions (SEDs) of blazars (e.g. \cite{GHISELLINI2010}), it is possible to estimate $k_{\rm CD}\sim 1$ for BL Lac Objects, and $k_{\rm CD}\sim 10$ for FSRQs. Therefore, we applied this correction to FSRQs and the results are displayed in Fig.~\ref{cdradpow}.

The comparison of the powers from 43~GHz and $\gamma-$ray observations does not change, with the linear fit giving these values: $m\sim 0.85$, $C\sim 6.3$, $\rho\sim 0.93$, $\sigma\sim 0.51$. However, the comparison with 15~GHz data is not so good: $m\sim 0.47$, $C\sim 22$, $\rho\sim 0.51$, $\sigma\sim 0.94$).

\begin{figure}[h]
\centering
\includegraphics[width=7 cm]{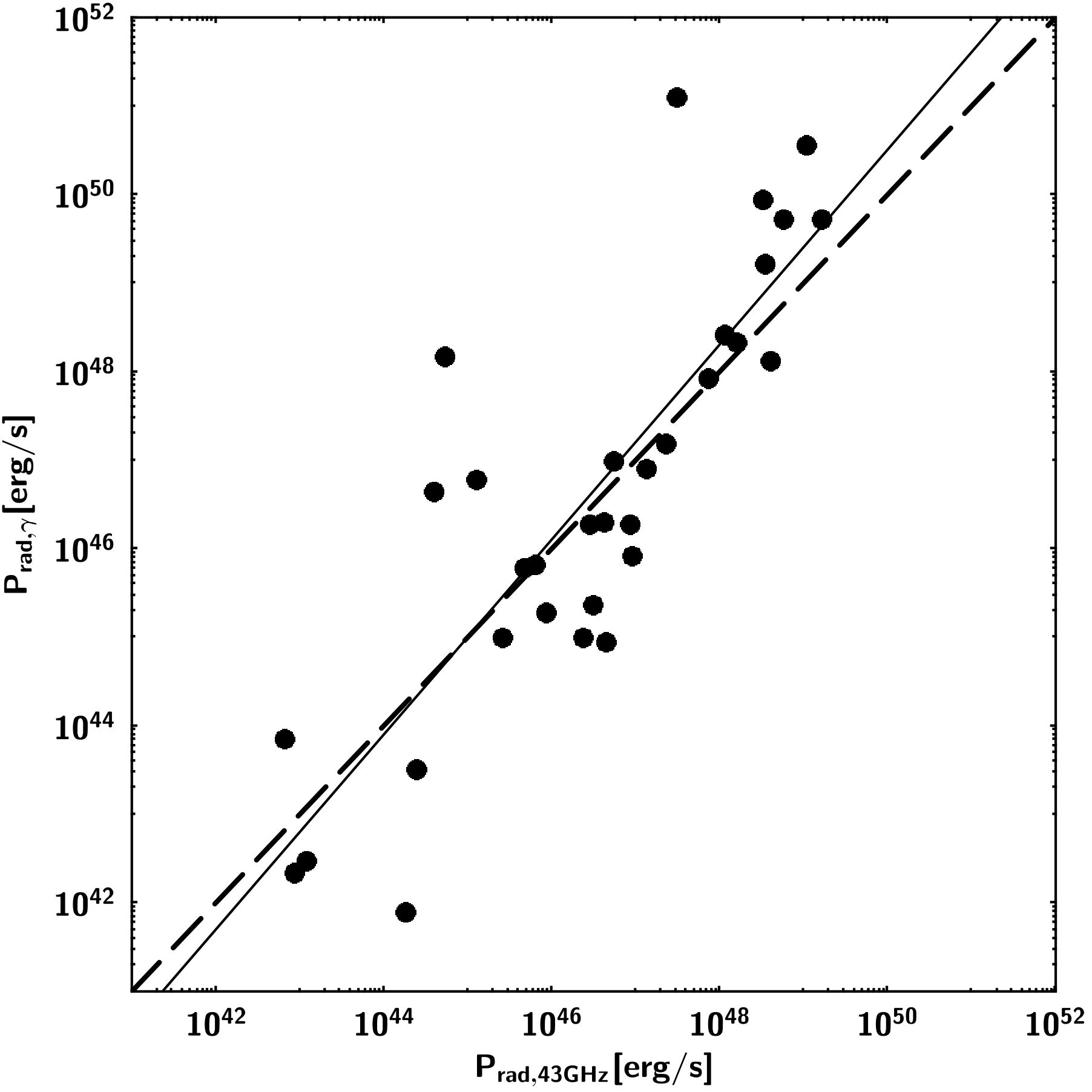}
\caption{\footnotesize Radiative jet power corrected for the Compton dominance. Comparison of values from high-energy $\gamma$ rays and radio observations at 43~GHz, with the Doppler factor calculated by using the brightness temperature (case 2, Sect.~\ref{VLBA}). The dashed line represents the equality of the two powers, while the continuous line is the linear fit to the data.  \label{TBradpow}}
\end{figure} 

The reason seems to be the use of the brightness temperature to estimate the Doppler factor, as shown already in Sect.~\ref{VLBA}. As a matter of fact, if we adopt the same method also for 43~GHz data, the consistency with the radiative power from $\gamma-$ray observations is lost (Fig.~\ref{TBradpow}). The linear fit is still acceptable, but with a large dispersion: $m\sim 1.10$, $C\sim -4.5$, $\rho\sim 0.83$, and $\sigma\sim 1.29$. Another source of bias is the use of a single value of $k_{\rm CD}$ for all FSRQs. This quantity depends on the characteristics of the source and its activity (an outburst can result in a greater value of $k_{\rm CD}$).

\section{Fudge factors}
\label{FF}
As noted in Sect.~\ref{VLBA}, the value of $k_{2}$ (see Eq.~\ref{fudge2}) is within a small range, particularly for 15~GHz data, with some exceptions. Therefore, we can try estimating the jet power by setting $k_2$ equal to a constant value (mean, median,...). We selected $k_2=0.183$, which is the median value calculated by selecting all the available epochs. Therefore, Eq.~(\ref{totaljetpow}) becomes: 

\begin{equation}
P_{\rm 44}\sim 4.5 \left(\frac{S_{\nu}d_{\rm L,9}^2}{1+z}\right)^{12/17}
\label{totaljetpow2}
\end{equation}

\begin{figure}[h]
\centering
\includegraphics[width=6.5 cm]{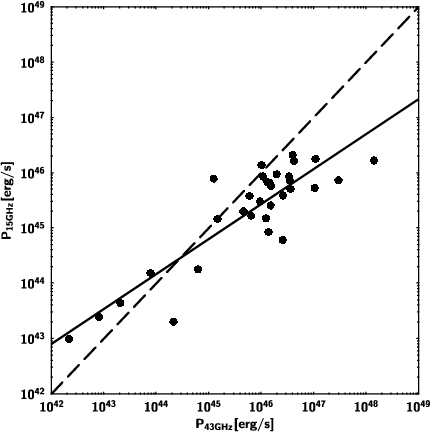}
\includegraphics[width=6.5 cm]{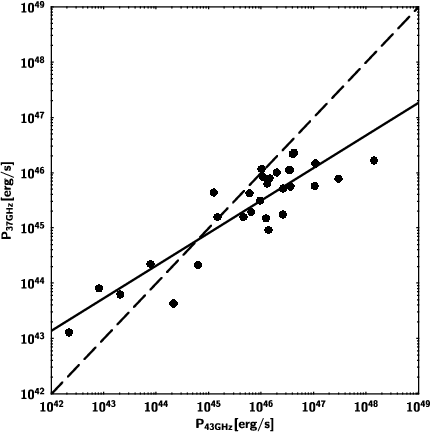}
\caption{\footnotesize Total jet power calculated with a constant $k_2$ (Eq.~\ref{totaljetpow2}) and flux densities at 15 ({\it left panel}) and 37~GHz ({\it right panel}), compared with the power at 43~GHz. The dashed line represents the equality of the two powers, while the continuous line is the linear fit to the data.  \label{constk2}}
\end{figure}

We then consider as reference the total jet power at 43~GHz, calculated with Eq.~(\ref{totaljetpow}), and compare it with the power at 15 and 37~GHz calculated with Eq.~(\ref{totaljetpow2}). The only variable is now the flux density at the selected frequency (15 or 37 GHz). The results are shown in Fig.~\ref{constk2}.

The linear fit gives the following results:

\begin{itemize}
\item	43 vs 15~GHz: $m\sim 0.63$, $C\sim 16$, $\rho\sim 0.87$, $\sigma\sim 0.44$;
\item	43 vs 37~GHz: $m\sim 0.59$, $C\sim 18$, $\rho\sim 0.89$, $\sigma\sim 0.38$;
\end{itemize}

\noindent with slightly better values for 37~GHz, as expected. However, the slope $\sim 0.6$ indicates a divergence toward low and high powers. We note that by selecting another value for $k_2$ (median or the average from another data set) will change only the value of $C$, but not all the others. The dispersion is contained within $\sim 0.4$.

\begin{figure}[h]
\centering
\includegraphics[width=6.5 cm]{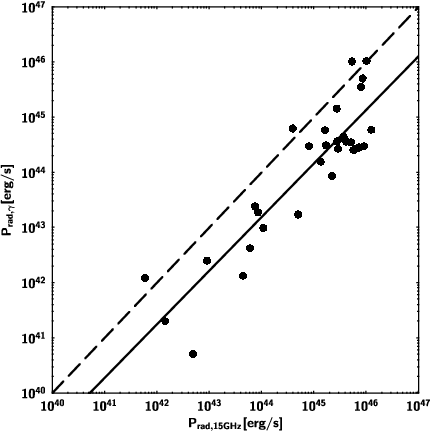}
\includegraphics[width=6.5 cm]{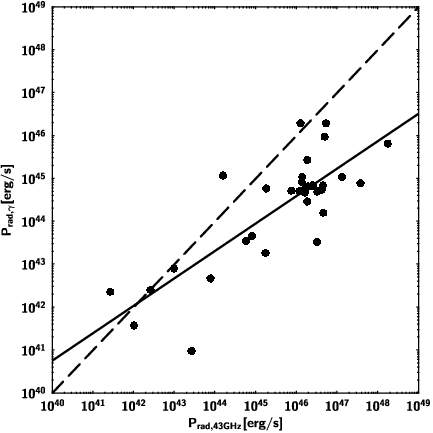}
\caption{\footnotesize Radiative jet power calculated with a constant $\Gamma^2/\delta^4$ compared with the power estimated with $\Gamma^2/\delta^4$ from 15~GHz ({\em left panel}) and 43~GHz data ({\em right panel}). The dashed line represents the equality of the two powers, while the continuous line is the linear fit to the data.  \label{fudgegamma}}
\end{figure}

We also studied the distributions of the correction factor $\Gamma^2/\delta^4$ to be applied to the $\gamma-$ray luminosity to estimate the radiative power (cf Eq.~\ref{radgamma}). We adopted the median calculated from all data, which is $\Gamma^2/\delta^4 \sim 0.0027$. We adopted the latter value as constant in Eq.~(\ref{radgamma}) and compared the radiative power estimated with the proper value for each source (Fig.~\ref{fudgegamma}). 

The result of the linear fit is now: 

\begin{itemize}
\item	$\gamma$ vs 15~GHz: $m\sim 0.97$, $C\sim 0.43$, $\rho\sim 0.90$, $\sigma\sim 0.56$;
\item	$\gamma$ vs 43~GHz: $m\sim 0.64$, $C\sim 15$, $\rho\sim 0.81$, $\sigma\sim 0.75$;
\end{itemize}

This time, there is a better agreement with 15~GHz data, but it is worth reminding that a good correlation does not imply a causation. This agreement is likely to be a chance coincidence, because the previous tests (see Sect.~\ref{LAT}, Fig.~\ref{radiativep}, left panel, and \ref{cdradpow}, left panel) do not display any hint of such agreement ($\rho\sim 0.51-0.54$). The only suitable explanation is that, by using constant fudge factors, most of fluctuations have been smoothed out by a mere chance coincidence. Taking constant average values for $k_2$ and $\Gamma^2/\delta^4$ has no physical reason and is only for our convenience to get rid of the lack of adequate measurements.

The comparison with the radiative power estimated from 43~GHz data is still acceptable, but with a larger dispersion and a divergence at high powers.

\section{Discussion and conclusions}
We compared the jet power as measured by different methods, mostly based on radio observations. We can summarise the main results as follows:

\begin{itemize}

\item The jet power estimates based on the Blandford \& K\"onigl model \cite{BLANDFORD1979} plus VLBA data at 15, and 43~GHz, are in good agreement (Sect.~\ref{VLBA}). The almost simultaneity of observations does not imply significant changes in the calculated jet power, at least with the present data set (Sect.~\ref{VLBA2}). One source of bias is the measurement of the Doppler factor $\delta$ via the brightness temperature (see Eq.~\ref{dopplertb}, and Fig.~\ref{dopplerBU2}). This problem has been already noted by several authors (e.g. \cite{JORSTAD2005,JORSTAD2017,LIODAKIS2018}, and particularly see the extensive discussion in \cite{HOMAN2021}), and is related to both the physics of the jets (opacity, absorption, activity of the jet,...) and the instrumental/observational issues (frequency, cadence of observations,...). We do not know the intrinsic brightness temperature for any source and cannot measure it. Therefore, we need either to make theoretical hypotheses \cite{READHEAD} or to follow a statistical approach, by assuming that every jetted AGN has more or less the same $T_{\rm b}$ equal to the median or the mean of the sample \cite{HOMAN2021}. The approach proposed by Jorstad et al. \cite{JORSTAD2005,JORSTAD2017} to calculate $\delta$ (cf Eq.~\ref{dopplerBU}) based on the flux variability is much more reliable, as shown by the excellent agreement with the radiative power measured from high-energy $\gamma-$rays (see Sect.~\ref{LAT}, particularly Fig.~\ref{cdradpow}, {\em right panel}). This approach seems to be not suitable for 15~GHz data, as radio observations at this frequency are sampling the jet downstream, where the flux variability is affected by effects other than radiative losses only \cite{HOMAN2021}. 

\item The use of single-dish flux densities at 37~GHz (Sect.~\ref{metsahovi}), with $k_2$ calculated from 15 and 43~GHz observations (see Eq.~\ref{fudge2}), is consistent with the power derived from VLBA observations. The best result is with 43~GHz data, as expected, because the smaller difference in frequency. 

\item The kinetic power calculated on the basis of the extended radio emission at MHz frequencies and the relationships by \cite{BIRZAN2008,CAVAGNOLO2010} (Sect.~\ref{kinetic}), gives better results when compared to the power estimated from the Blandford \& K\"onigl \cite{BLANDFORD1979} model and 15~GHz data. However, we noted a systematic disagreement of the power for weak sources ($P_{\rm kin}\lesssim 10^{44}$~erg/s). 

\item The comparison of the radiative power estimated from the Blandford \& K\"onigl \cite{BLANDFORD1979} model and high-energy $\gamma-$ray observations from {\em Fermi}/LAT (Sect.~\ref{LAT}) resulted in an excellent agreement, particularly with 43~GHz data, and when taking into account the Compton dominance. The larger dispersion in the comparison with 15~GHz data seems to be due to the above cited limitations of $\delta$ calculated via $T_{\rm b}$ (Fig.~\ref{TBradpow}). However, a quite good agreement with 15~GHz data is recovered when using a constant value for $\Gamma^2/\delta^4$ to estimate the radiative power, even though it is systematically lower than the value from radio observations and is likely to be a chance coincidence (Sect.~\ref{FF}).

\item Searching for an easy-to-use equation to estimate the jet power, we proposed Eq.~(\ref{totaljetpow2}), based on the limited range of values of $k_{2}$, particularly from 15~GHz data. The comparison of power derived from 15, 37, and 43~GHz data are fairly correlated ($\rho \sim 0.9$) with an acceptable dispersion $\sigma \sim 0.4$. The use of a constant $\Gamma^2/\delta^4$ to estimate the radiative power from the $\gamma-$ray luminosity resulted in a slightly greater dispersion ($\sigma\sim 0.6-0.7$). 

\end{itemize}

For the sake of simplicity, we recall in Table~\ref{easypower} the proposed easy-to-use equations to estimate the jet power, with the caveat of divergence at low and high powers. 

\begin{table}[h!] 
\caption{\footnotesize Jet power in [erg~s$^{-1}$] calculated with our proposed easy-to-use equations based on fudge factors described in Sect.~\ref{FF}. We remind that the radio flux density $S_{\nu}$ is measured in [Jy], the luminosity distance $d_{\rm L,9}$ is in [Gpc], and $L_{\gamma}$ is in [erg~s$^{-1}$].  \label{easypower}}
\centering
\vskip 6pt
\begin{tabular}{lcl}
\hline
\textbf{Jet Power}	& \textbf{Equation}	& \textbf{Notes}\\
\hline
Total		& 	$(4.5\times 10^{44}) \left(\frac{S_{\nu}d_{\rm L,9}^2}{1+z}\right)^{\frac{12}{17}}$		& From Eq.~(\ref{totaljetpow})\\
Kinetic 	& $(3.9\times 10^{44}) \left(\frac{S_{\nu}d_{\rm L,9}^2}{1+z}\right)^{\frac{12}{17}}$			&  From Eq.~(\ref{kinpow})\\
Radiative (synchrotron)		& $(5.6\times 10^{43}) \left(\frac{S_{\nu}d_{\rm L,9}^2}{1+z}\right)^{\frac{12}{17}}$			& From Eq.~(\ref{synchropow}).\\
Radiative (Compton)		& $0.0027 L_{\gamma}$			& From Eq.~(\ref{radgamma})  \\
\hline
\end{tabular}
\end{table}

We want to stress that equations in Table~\ref{easypower} must be used with great care, because the fudge factors are affected by the variability of the source and the uncertainties in the measurement or derivation of the physical quantities $\beta,\Gamma,\delta,\theta$, and $\phi$ (that we did not consider in this work). However, given the difficulty to measure or to infer all these quantities without dedicated VLBA observations (preferably at high frequencies, such as 43 GHz), these equations can offer a useful first estimate of the jet power, and being careful when dealing with extremely weak or extremely powerful jets.

Before concluding, some more words of caveat should be written, which are also the points to be addressed to improve our methods to estimate the jet power. The possible sources of bias in the present work are:

\begin{itemize}
\item the sample is composed mostly of blazars (30/32 objects), whose electromagnetic emission is dominated by relativistic beaming, because of the small viewing angle. Only two objects are misaligned AGN (radio galaxies), and there are no jetted Seyferts. It is necessary to expand the sample to cover all the types of jetted AGN, beamed or not.

\item to convert redshifts into luminosity distances, we employed the simplified Eq.~(\ref{distance}). This resulted in an overestimation of the luminosity distance of $\sim 10$\% for the farthest object (J$0841+7053$, $z=2.71$), which quickly decreases to $\sim 4$\% for objects at $z\sim 1$. This is not a problem in the present work, since we compared the jet power of the same object calculated with different methods, but the comparison with values from other works should be dealt with care in the case of high-redshift objects. 

\item the Blandford \& K\"onigl \cite{BLANDFORD1979} model is for flat-spectrum radio sources. Deviation from a flat radio spectrum, such as in the cases of steep spectra of misaligned AGN, might imply large errors. In our sample, we have only two radio galaxies, too few to draw useful conclusions;

\item the extended radio emission to estimate the kinetic power (Sect.~\ref{kinetic}) should be only due to radio lobes, with a steep spectrum. However, for the sake of simplicity, we considered the whole integrated flux. As a matter of fact, the typical resolution at $200-400$~MHz is about one arcminute, which is equivalent to $\sim 0.1$~Mpc at $z\sim 0.1$. Therefore, most of the objects in our sample are point-like at MHz frequencies, and it is not possible to isolate the steep spectrum extended emission from the core. Anyway, at MHz frequencies the core contribution should be less important than the lobes. The Low Frequency Array (LOFAR) might be a viable solution for a better angular resolution ($\sim 0.21''$ at 240~MHz for a 1000~km baseline\footnote{\url{https://science.astron.nl/telescopes/lofar/lofar-system-overview/observing-modes/lofar-imaging-capabilities-and-sensitivity/}}), but it is necessary to recalibrate the Eqs.~(\ref{birzan}) and (\ref{cavagnolo}), because the maximum frequency of LOFAR is 250~MHz.   

\item in this work, we always used median or weighted mean values calculated over long periods. The shortest period is $2007-2013$, about 5.5 years. Given the strong variability of jetted AGN, the use of values from single-epoch observations or from only one VLBA knot might result in significant deviations. For example, we considered J$0433+0521$ with VLBA data at 43~GHz: the total jet power with the data used in this work results to be $\sim 2.1\times 10^{44}$~erg/s. We want to compare with the most recent data from \cite{WEAVER2022}, who extended the work by \cite{JORSTAD2017} to 2018 December. By using the median values, we calculate $\sim 4.1\times 10^{44}$~erg/s, consistent within a factor 2 with the present work. If we calculate the jet power by using the data -- for example -- of the component C15 only, we obtain $\sim 5.2\times 10^{43}$~erg/s, about one order of magnitude smaller. 

\item we need also to underline that this work was done by considering the same physical factors $\Delta=\log (r_{\rm max}/r_{\rm min})$, and $\Lambda=\log (\gamma_{\rm e,max}/\gamma_{\rm e,min})$ for all the sources. Therefore, a part of the dispersions in the comparisons is surely due to this assumption. For example, an outburst changing the electron distribution will alter $\Lambda$, which in turn change the coefficient $k_1$ of Eq.~(\ref{fudge1}). Therefore, it is necessary to address also the microphysics of the jet, and, particularly, the particle content (leptons vs hadrons), the energy distribution of electrons, the size of the emission region vs opacity, the equipartition hypothesis.

\end{itemize}

\vspace{6pt} 
\section*{Acknowledgments}
LF would like to thank (in alphabetical order) Stefano Ciroi, Maria~J.~M. March\~{a}, Patrizia Romano, and Stefano Vercellone for valuable comments on the draft. This research made use of publicly available data from the VLBA Boston University blazar program \cite{BUBLAZAR,JORSTAD2017}, the MOJAVE program \cite{LISTER2018}, the Mets\"ahovi Radio Observatory\footnote{\url{https://www.metsahovi.fi/opendata/}}, the CATS database \cite{CATS}, and the {\em Fermi} LAT Light Curve Repository \cite{LATLCR}. Recalculated or new values resulted from this work are available upon reasonable requests.

\end{document}